\documentclass[10pt,twocolumn,letterpaper]{article}
\usepackage[usenames, dvipsnames]{xcolor}
\usepackage{usenix2019, times,color,graphicx,amsmath,array,amssymb, subcaption}
\usepackage{xspace}

\newcommand{\name}{{\sc{GEMEL}}\xspace}

\newcommand{\PAGENUMBERS}{yes}
\newcommand{\COMMENTS}{yes}

\usepackage{cite}
\usepackage[T1,hyphens]{url}

\usepackage[
  colorlinks=true,
  citecolor=red
]{hyperref}

\def\moreskip{\vspace\moreskipamount}
\newskip\moreskipamount \moreskipamount=2pt plus 1pt minus 1pt

\newcommand{\ie}{{i.e.,}~}
\newcommand{\eg}{{e.g.,}~}

\newcommand{\para}[1]{\moreskip\noindent\textbf{#1}}

\usepackage{balance, ifthen, multirow}

\usepackage[small,compact]{titlesec}              

\usepackage[font=small,labelfont=bf,textfont=bf,aboveskip=8pt]{caption}
\usepackage{subcaption}

\usepackage{fancyhdr}

\ifthenelse{\equal{\PAGENUMBERS}{yes}}{%
  \pagestyle{plain}
}{%
  \pagestyle{empty}
}

\usepackage{enumitem}
\setlist{itemsep=0pt,parsep=0pt,topsep=0pt}

\usepackage{booktabs}
\usepackage{colortbl}
\usepackage{float}                           

\definecolor{placeholderbg}{rgb}{0.85,0.85,0.85}

\usepackage{mathtools}
\usepackage{pifont}

\usepackage{url}
\usepackage{cleveref}
\crefname{section}{\S}{\SS}
\crefformat{section}{\S#2#1#3} 
\crefformat{subsection}{\S#2#1#3}
\crefformat{subsubsection}{\S#2#1#3}

\usepackage{listings}
\newcommand\code[1]{\lstinline$#1$}

\lstdefinelanguage{paper}{
 keywords={partition, transform, gather, scatter, apply},
 keywordstyle=\color{blue}\bfseries,
 morekeywords={[2]degrees,branch,commit,v_prev},
 keywordstyle={[2]\color{red}\bfseries},
 morekeywords={[3]if,def,Class,return,else,None,False,True,Array,while,G},
 keywordstyle={[3]\bfseries},
 basicstyle=\small\ttfamily,
 identifierstyle=\color{black},
 sensitive=false,
 comment=[l]{\/\/},
 morecomment=[s]{/*}{*/},
 commentstyle=\color{green}\ttfamily,
 stringstyle=\color{red}\ttfamily,
 breaklines=true,
}
\newcommand{\tightcaption}[1]{\vspace{-11pt}\caption{{\bf \small #1}}
\vspace{-15pt}
}
\lstnewenvironment{python}
{\lstset{
    language=paper,
    frame=tlbr,
    columns=fullflexible,
    postbreak=\mbox{\textcolor{red}{$\hookrightarrow$}\space},
    captionpos=b}}
{}

\newcommand{\squishlist}{
   \begin{list}{$\bullet$}
    { \setlength{\itemsep}{0pt}      \setlength{\parsep}{3pt}
      \setlength{\topsep}{1pt}       \setlength{\partopsep}{0pt}
      \setlength{\leftmargin}{1.0em} \setlength{\labelwidth}{1em}
      \setlength{\labelsep}{0.5em} } }
\newcommand{\squishend}{
    \end{list}  }

\usepackage{algorithm}
\usepackage{algpseudocode}

\usepackage{tikz}
\DeclareRobustCommand\numcircledtikz[1]{\tikz[baseline=(char.base)]{
    \node[shape=circle,draw,fill,inner sep=1pt] (char)
    {\textcolor{white}{#1}};}}

\ifthenelse{\equal{\COMMENTS}{yes}}{%
    \setlength{\marginparwidth}{1.5cm}
    \usepackage[draft]{todonotes}
}{%
    \usepackage[disable]{todonotes}
}%

\newcommand{\add}[1] {{\textcolor{black}{{#1}}}}
\newcommand{\nop}[1] {{{{}}}}

\ifthenelse{\equal{\COMMENTS}{yes}}{%
  \newcommand{\ap}[1] {{\textcolor{magenta}{AP: {#1}}}}
    
  \newcommand{\rn}[1] {{\textcolor{red}{RN: {#1}}}}
  \newcommand{\fillin}[1] {{{{#1}}}}

  \newcommand{\hx}[1] {{\textcolor{purple}{HX: {#1}}}}
  \newcommand{\ga}[1] {{\textcolor{orange}{GA: {#1}}}}
  \newcommand{\ys}[1] {{\textcolor{green}{YS: {#1}}}}
}{%
  \newcommand{\ap}[1] {}
  \newcommand{\rn}[1] {}
  \newcommand{\hx}[1] {}
  \newcommand{\ga}[1] {}
  \newcommand{\ys}[1] {}
}%

\usepackage{comment}
\usepackage{soul}
\soulregister\cite7
\soulregister\ref7
\soulregister\pageref7

\usepackage[all]{nowidow}

\begin{document}

\twocolumn[\begin{@twocolumnfalse}

\begin{centering}

{\large \bf \name{}: Model Merging for Memory-Efficient, Real-Time Video Analytics at the Edge \vspace{-0.1cm}}

\def\refUCLA{$^\mathsection$}
\def\refPrinceton{$^\P$}
\def\refmsr{$^\dagger$}

{\vspace{0.5cm}
\begin{tabular}[t]{cccc}
Arthi Padmanabhan$^\star$\refUCLA{}&Neil Agarwal$^\star$\refPrinceton{}&Anand Iyer\refmsr{}&Ganesh Ananthanarayanan\refmsr{}\\
Yuanchao Shu\refmsr{}&Nikolaos Karianakis\refmsr{}&Guoqing Harry Xu\refUCLA{}&Ravi Netravali\refPrinceton{}\\
\end{tabular}\par}

{\vspace{0.2cm}
\begin{tabular}[t]{ccc}
\refUCLA{}UCLA&\refmsr{}Microsoft Research&\refPrinceton{}Princeton University \\
\end{tabular}\par}

\end{centering}

\vspace{\baselineskip}

\end{@twocolumnfalse}]
\begin{abstract}
Video analytics pipelines have steadily shifted to edge deployments to reduce bandwidth overheads and privacy violations, but in doing so, face an ever-growing resource tension. Most notably, edge-box GPUs lack the memory needed to concurrently house the growing number of (increasingly complex) models for real-time inference. Unfortunately, existing solutions that rely on time/space sharing of GPU resources are insufficient as the required swapping delays result in unacceptable frame drops and accuracy loss. We present \emph{model merging}, a new memory management technique that exploits architectural similarities between edge vision models by judiciously sharing their layers (including weights) to reduce workload memory costs and swapping delays. Our system, \name{}, efficiently integrates merging into existing pipelines by (1) leveraging several guiding observations about per-model memory usage and inter-layer dependencies to quickly identify fruitful and accuracy-preserving merging configurations, and (2) altering edge inference schedules to maximize merging benefits. Experiments across diverse workloads reveal that \name{} reduces memory usage by up to \fillin{60.7}\%, and improves overall accuracy by \fillin{8-39}\% relative to time or space sharing alone.
\end{abstract}

\ifthenelse{\equal{\PAGENUMBERS}{no}}{%
  \thispagestyle{empty}
}

\makeatletter
\def\blfootnote{\xdef\@thefnmark{}\@footnotetext}
\makeatother

\blfootnote{$^\star$ These authors contributed equally to this work.}

\section{Introduction}\label{s:intro}

Fueled by the proliferation of camera deployments and significant advances in deep neural networks (DNNs) for vision processing (e.g., classification, detection)~\cite{cnn-face-cvpr15, pyramid-network-cvpr17, pedestrian-detection-iccv15, maskrcnn,pytorchyolov3}, live video analytics have rapidly grown in popularity~\cite{dds,reducto,EdgeSurvey2,VideoKillerApp2,chameleon}. Major cities and organizations around the world now employ thousands of cameras to monitor intersections, homes, retail spaces, factories, and more~\cite{LondonCamera,paris-hospital, beijing-cameras,ChicagoCamera}. The generated video feeds are continuously and automatically queried using DNNs to power long-running applications for autonomous driving, footfall tracking, traffic coordination, business analytics, and surveillance~\cite{are-we-ready-for-ai-powered-security-cameras,powering-the-edge-with-ai-in-an-iot-world,vision-zero,smart-mall,traff2}.

In order to  deliver highly-accurate query responses in real time, video analytics deployments have steadily migrated to the edge~\cite{VideoKillerApp2,edgenative,vidband}.
More specifically, pipelines routinely incorporate \emph{on-premise} edge servers (e.g., Microsoft Azure Stack Edge~\cite{azure_stack_edge}, Amazon Outposts~\cite{aws_outposts}) that run in hyper-proximity to cameras (in contrast to traditional edge servers~\cite{akamai,googleinfra,mysterymachine,fogdef}), and possess on-board GPUs to aid video processing. These \emph{edge boxes} are used to complement (or even replace~\cite{ekya,filterforward}) distant cloud servers by locally performing as many inference tasks on live video streams as possible~\cite{videoedge,reducto,vigil,filterforward}. Generating responses directly on edge boxes reduces transfer delays for shipping data-dense video over wireless links~\cite{Han:2016:MAV:2999572.2999606,vigil,spider} while also bringing resilience to outbound edge-network link failures~\cite{chicfila,ofcom} and compliance with regional data privacy restrictions~\cite{microsoftprivacy,visor}.

\begin{figure}[!tbp]
\center
    \includegraphics[width=0.9\linewidth]{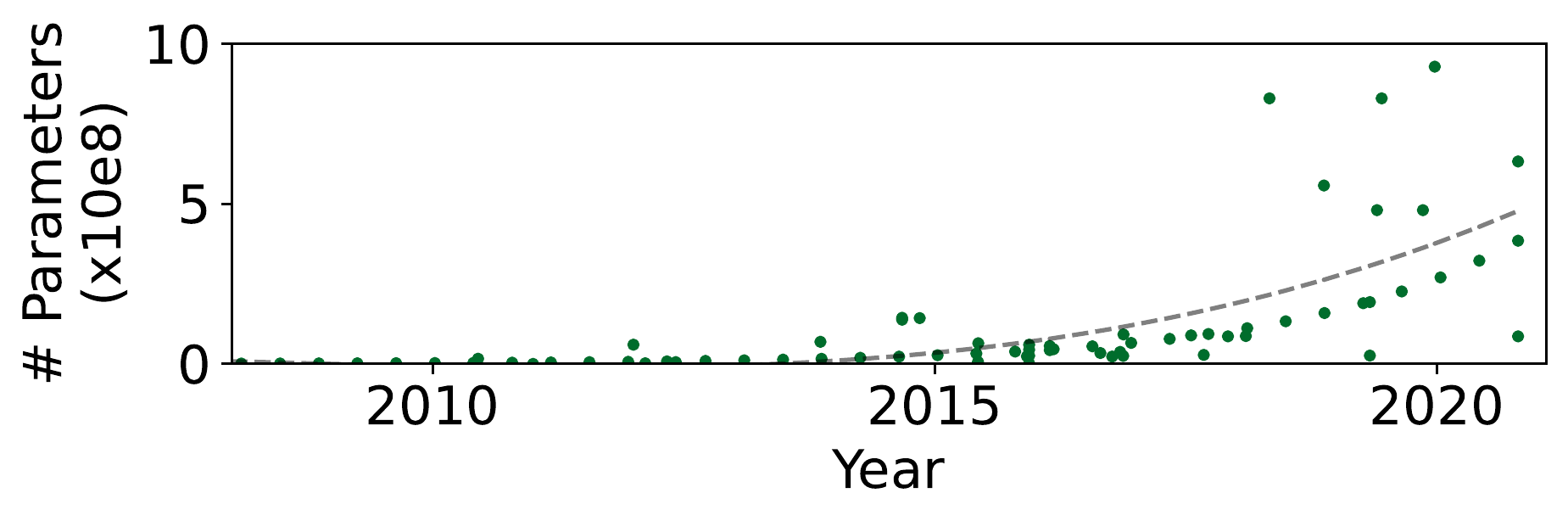}
    \vspace{-12pt}
    \caption{Parameter counts in popular vision DNNs over time. Data drawn from~\cite{visparams}.}
    \vspace{-3pt}
    \label{fig:memory_over_time}
  \end{figure}

To reap the above benefits, video analytics deployments must operate under the limited computation resources offered by edge boxes. On the one hand, due to cost, power, and space constraints, edge boxes typically possess weaker GPUs than their cloud counterparts~\cite{ekya,shi2016edge,azure_stack_edge}.
On the other hand, analytics deployments face rapidly increasing workloads due to the following trends: (1) more camera feeds to analyze~\cite{videoedge,ekya,spatula}, (2) more models to run due to increased popularity and shifts to bring-your-own-model platforms~\cite{rocket,amazon-rekognition,google-cloud-vision,ibm-maximo}, and (3) increased model complexity, primarily through growing numbers of layers and parameters (Figure~\ref{fig:memory_over_time})~\cite{EdgeMemoryBound,DNNWorkloadAnalysisATC19, DNNWorkloadAnalysisCoRR, DNNWorkloadAnalysisTR}. Taken together, the result is an ever-worsening resource picture for edge video analytics. 

\para{Problems.} Although GPU computation resources are holistically constrained on edge boxes, this paper focuses on \emph{GPU memory restrictions}, which have become a primary bottleneck in edge video analytics for three main reasons. First, GPU memory is costly due to its high-bandwidth nature~\cite{monet,zeroinf,capuchin}, and is thus unlikely to keep pace with the ever-growing memory needs of DNNs (Figure~\ref{fig:memory_over_time}). Second, we empirically find that existing memory management techniques that time/space-share GPU resources~\cite{SwapAdvisor, nexus, antman, pipeswitch, clockwork,DNNWorkloadAnalysisATC19} are insufficient for edge video analytics, resulting in skipped processing on \fillin{19-84}\% of frames, and corresponding accuracy drops up to \fillin{43\%} (\S\ref{sec:motivation}). The underlying reason is that the costs of loading vision DNNs into GPU memory (i.e., swapping) are prohibitive and often exceed the corresponding inference times, leading to sub-frame-rate ($<$ 30 fps) processing and dropped frames due to SLA violations~\cite{nexus,deadline1}. Such accuracy drops are unacceptable for important vision tasks, especially given that each generation of vision DNNs brings only 2-10\% of accuracy boosts -- that after painstaking tuning~\cite{visionsurvey1,visionsurvey2,visionsurvey3,mediumvis}. Third, compared to computation bottlenecks~\cite{chameleon,reducto,clockwork,filterforward,mistify}, GPU memory restrictions during inference have been far less explored in video analytics. 

\para{Contributions.} We tackle this memory challenge by making two main contributions described below. The design and evaluation of our solution are based on a wide range of popular vision DNNs, tasks, videos, and resource settings that reflect workloads observed in both our own multi-city pilot video analytics deployment and in prior studies (\S\ref{s:workloads}).

\underline{Our first contribution is \emph{model merging}}, a fundamentally new approach to tackling GPU memory bottlenecks in edge video analytics that is complementary to time/space-sharing strategies (\S\ref{s:merging}). With merging, we aim to share \emph{architecturally identical} layers across the models in a workload such that only one copy of each shared layer (\ie one set of weights) must be loaded into GPU memory for all models that include it. In doing so, merging reduces both the number of swaps required to run a workload (by reducing the overall memory footprint) and the cost of each swap (by lowering the amount of new data to load into GPU memory).

 Our merging approach is motivated by our (surprising) finding that vision DNNs share substantial numbers of layers that are architecturally (\ie excluding weights) identical (\S\ref{ss:potential}). Such commonalities arise not only between identical models (100\% sharing), but also across model variants in the same (up to \fillin{\add{84.6\%}}) and in different (up to \fillin{\add{96.3}}\%) families. The reason is that, despite their (potentially) different goals, vision DNNs ultimately employ traditional CV operations (\eg convolutions)~\cite{visionsurvey1,visionsurvey2}, operate on unified input formats (\eg raw frames), and perform object-centric tasks (\eg detection, classification) that rely on common features such as edges, corners, and motion~\cite{background-subtraction-cvpr2011,event-camera-eccv16,keyframe-odometry-bmvc17,slam-ral18,inertial-odometry-cvpr17,feature-tracks-iros16,probabilistic-data-icra17, glimpse-sensys15}.

Our analysis reveals that exploiting these architectural commonalities via merging has the potential to substantially lower memory usage (\fillin{17.9-86.4}\%) and boost accuracy (\fillin{by up to 50}\%) in practice. However, achieving those benefits is complicated by the fact that edge vision models typically use different weights for common layers due to training non-linearities~\cite{nonlin1,nonlin2} and variance in target tasks, objects, and videos; and yet, merging requires using unified weights for each shared layer. Digging deeper, we observe that there exists an \emph{inverse relationship} between the number of shared layers and achieved accuracy during retraining. Intuitively, this is because for shared layers to use unified weights, other layers must adjust their weights accordingly during retraining; the more layers shared, the harder it is for (the fewer) other layers to find weights to accommodate such constraints and successfully learn the target functions~\cite{overparam1,overparam2}. Worse, determining the right layers to merge is further complicated by the fact that (1) it is difficult to predict precisely how many layers will be shareable before accuracy violations occur, and (2) each instance of retraining is costly.

\underline{Our second contribution is \name{}}, 
an end-to-end system that practically incorporates model merging into edge video analytics by automatically finding and exploiting merging opportunities across user-registered vision DNNs (\S\ref{s:gemel}). \name{} tackles the above challenges by leveraging two key observations: (1) vision DNNs routinely exhibit power-law distributions whereby a small percentage of layers, often towards the end of a model, account for most of the model's memory usage, and (2) merging decisions are agnostic to inter-layer dependencies, and accordingly, a layer's mergeability does not improve if other layers are also shared.

Building on these observations, \name{} follows an \emph{incremental} merging process whereby it attempts to share one additional layer during each iteration, and selects new layers in a memory-forward manner, \ie prioritizing the (few) memory-heavy layers. In essence, this approach aims to reap most of the potential memory savings as quickly, and with as few shared layers, as possible. \name{} further accelerates the merging process by taking an adaptive approach to retraining that detects and leverages signs of early successes and failures. At the end of each successful iteration, \name{} ships the resulting merged models to the appropriate edge servers, and carefully alters the time/space-sharing scheduler -- a merging-aware variant of Nexus~\cite{nexus} in our implementation -- to maximize merging benefits, \ie by organizing merged models to reduce the number of swaps, and the delay for each one. Importantly, \name{} verifies that merging configurations meet accuracy targets \emph{prior} to deployment at the edge, and also periodically tracks data drift.

\paragraph{Results.} We evaluated \name{} on a wide range of workloads and edge settings (\S\ref{s:workloads}, \S\ref{ss:generalization}). Overall, \name{} reduces memory requirements by up to \fillin{60.7\%}, which is \fillin{5.9-52.3}\% more than stem-sharing approaches that are fundamentally restricted to sharing contiguous layers from the start of models (Mainstream~\cite{MainstreamATC2018}), and within \fillin{9.3-29.0}\% of the theoretical maximum savings (that disregard layer weights). These memory savings lead to \fillin{13-44}\% fewer skipped frames and overall accuracy improvements of \fillin{8-39}\% compared to space/time-sharing GPU schedulers alone (Nexus~\cite{nexus}). We will open-source \name{} and our datasets.

\section{Methodology \& Pilot Study}
\label{s:workloads}

We begin by describing the workloads used in this paper. They were largely derived from our experience in deploying a pilot video analytics system in collaboration with two major US cities (one per coast), for road traffic monitoring.

\para{Models and tasks.} \add{In line with other video analytics frameworks~\cite{rocket,amazon-rekognition, google-cloud-vision, ibm-maximo}, users in our deployment provided pre-trained models when registering queries to run on different video feeds. Due to the complexity of model development, we observe that users opt to leverage existing (popular) architectures geared for their target task (e.g., YOLOv3 for object detection), and train those models for specific object(s) of interest and datasets (\eg detecting vehicles at Main St.) to generate a unique set of weights. Despite being allowed, custom architectures were never provided in our deployment.}

Accordingly, we selected the 7 most popular families of models across our pilot deployment and recent literature~\cite{NoScopeVLDB2017,FocusOSDI2018,MainstreamATC2018,chameleon,reducto,cloudseg,pipeswitch,ekya,SwapAdvisor,videoedge}: YOLO, Faster RCNN, ResNet, VGG, SSD, Inception, and Mobilenet. From each family,
we selected up to 4 model variants (if available) that exhibit different degrees
of complexity and compression. For instance, from YOLO, we consider
\{YOLOv3, Tiny YOLOv3\}; similarly, we consider
ResNet\{18, 50, 101, 152\}. The selected models focus on two tasks -- object classification and detection -- and for each, we train different versions for \add{all combinations of the following objects: people and vehicles (e.g., cars, trucks, motorbikes).}
Classification and detection
accuracy are measured using F1 and mAP~\cite{Everingham:2010:PVO:1747084.1747104}. 

\para{Videos.} Our dataset consists of video streams from 12 cameras in our pilot deployment that span two metropolitan areas. From each region, we consider cameras at adjacent intersections, and those spaced farther apart
within the same metropolitan area; this enables us to consider different edge box placements, i.e., at a traffic intersection vs.  further upstream to service a slightly larger geographic location.
From each stream, we scraped 120 minutes of video that cover 24-hour periods from four times of the year. 

\para{Edge boxes.} Our review of on-premise edge boxes focused on 5
commercial offerings: Microsoft Azure Stack Edge~\cite{azure_stack_edge}, Amazon
Outposts~\cite{aws_outposts}, Sony REA~\cite{sony}, NVIDIA
Jetson~\cite{nvidia_jetson}, and Hailo Edge-AI-box~\cite{hailo}. These servers
each possess on-board GPUs and offer 2-16 GB of total GPU memory. Since
edge inferences do not typically span multiple GPUs, we focus on model merging
and inference scheduling \emph{per GPU}. This does not
restrict \name{} to single-GPU settings; rather, it means that our merging and
scheduling techniques are applied separately to the DNNs in each GPU, with
the assumption that each merged model runs on only one GPU.

\para{Workload construction.} Recent works highlight that 10s of videos
are usually routed to each edge box~\cite{videoedge,vision-zero},
which runs upwards of 10 queries (or DNNs) on each feed~\cite{ekya,rocket}. Our
experience was similar: it was typical to direct the max possible number of feeds to an edge box, with the goal of \emph{minimizing the number of edge boxes required to process the workload}. To cover this space, and since we focus on per-GPU inference optimization, we generated an exhaustive list of all possible workloads sized between \add{2-50} DNNs using the models above. We then sorted the list in terms of the potential \add{(percentage)} memory savings (using the methodology from \S\ref{s:merging}), and selected 15 workloads: 3 random workloads from the lower quartile (i.e., \emph{Low Potential (LP1-3)}), 6 from the middle 50\% (i.e., \emph{Medium Potential (MP1-6)}), and 6 from the upper quartile (i.e., \emph{High Potential (HP1-6)}). \add{We chose this ratio to match that from our deployment. MP and HP workloads each constituted 30-50\% of the total workloads since (1) users
tended to employ the same few model variants from a limited set of popular families, and (2) each user typically used the same architecture (but not weights) for different feeds in a region. LP workloads were less common ($<$20\%), and arose from different users opting for different model families.}

Each workload was randomly assigned to one of the cities, with the constituent models being randomly paired with the available videos. \S\ref{ss:workload_details} details the workloads, each of which exhibits heterogeneity in terms of the families, tasks, videos, and (combinations of) target objects of the included models. \add{In summary, the workloads contain 3-42 queries (avg: 15) across 3-7 video feeds (avg: 5), featuring 2-10 unique models (avg: 6) and 2-5 different objects (avg: 4).}
\add{We consider additional workloads, models, objects, and videos in \S\ref{ss:generalization}.}

\para{Result presentation.} End-to-end accuracy depends on the available GPU memory. However, each workload requires a different minimum amount of memory to run, \ie the GPU should be able to load/run the most memory-intensive model in isolation for a batch size of 1. Further, the memory needed to avoid swapping (\ie to load all models and run one at a time) also varies per workload; we call this no\_swap. To ensure comparability across all presented accuracy results and to focus on memory-bottlenecked scenarios, we assign each workload three memory settings to be evaluated on (listed in \S\ref{ss:memory_settings}): (1) the minimum value (\emph{min}), (2) 50\% of the no\_swap value (\emph{50\%}), and (3) 75\% of the no\_swap value (\emph{75\%}).
\section{Motivation}
\label{sec:motivation}

\begin{figure}[!tbp]
\center
    \includegraphics[width=0.77\linewidth]{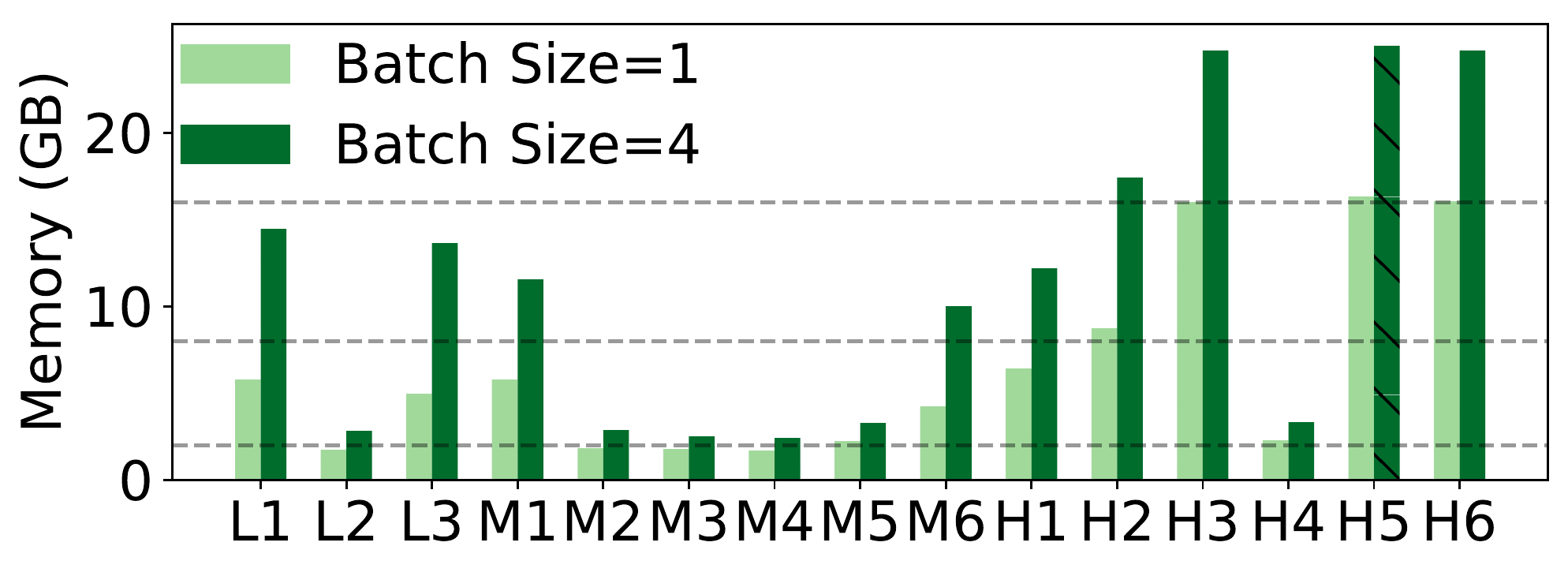}
    \vspace{-10pt}
    \caption{Per-workload memory requirements for two popular batch sizes used in video analytics~\cite{nexus}. Dashed lines represent the available GPU memory on several commercial edge boxes.}
    \vspace{-1pt}
    \label{fig:memory_fit}
  \end{figure}

\subsection{Memory Pressure in Edge Video Analytics}
\label{ss:problem}

To run inference with a given model, that model's layers and parameters must be loaded into the GPU's memory, with sufficient space reserved to house intermediate data generated while running, \eg activations. The amount of data generated (and thus, memory consumed) during inference depends on both the model architecture and the batch size used; a higher batch size typically requires more memory.

Figure~\ref{fig:memory_fit} shows the total amount of memory (i.e., for both loading and running)  required for each of our workloads and two batch sizes; the listed numbers exclude the fixed memory that ML frameworks reserve for operation, e.g., 0.8 GB for PyTorch~\cite{PyTorch}. As shown, many workloads do not directly fit into edge box GPUs, and the number of edge boxes necessary to support a given workload can quickly escalate. For instance, even with a batch size of 1 frame, 73\% of our workloads need more than one edge box possessing 2 GB of GPU memory; with a batch size of 4, 60\% and 27\% require more than one edge box with 8 GB and 16 GB of memory.

\begin{table}[t]
\scriptsize
\centering
\begin{tabular}{@{}lllll@{}}
\toprule
\multicolumn{1}{l}{\multirow{2}{*}{\textbf{Model}}} & \multirow{2}{*}{\textbf{Load Memory}} & \multicolumn{3}{c}{\textbf{Run Memory (Time)}} \\ \cmidrule(l){3-5}
\multicolumn{1}{l}{}             &    \textbf{(Time)}   & \textbf{BS=1} & \textbf{BS=2} & \textbf{BS=4} \\ \midrule
YOLOv3                           & 0.24 (49.5)  & 0.52 (17.0)      & 0.73 (24.0)    & 1.22 (39.9)          \\
ResNet152                        & 0.24 (73.3)  & 0.65 (24.8)      & 0.98 (26.3)    & 1.71 (26.7)         \\
ResNet50                         & 0.12 (27.1)  & 0.35 (8.4)       & 0.50 (8.5)     & 0.84 (8.5)        \\
VGG16                            & 0.54 (72.2)  & 0.74 (2.1)       & 0.89 (2.4)     & 1.18 (2.4)        \\
Tiny YOLOv3                      & 0.04 (6.7)   & 0.15 (3.0)       & 0.18 (5.2)     & 0.24 (5.2)        \\
Faster RCNN                      & 0.73 (117.3) & 3.70 (115.4)     & 6.96 (210.1)   & 12.47 (379.4)         \\
Inceptionv3                      & 0.12 (11.8)  & 0.19 (9.1)       & 0.23 (9.1)     & 0.34 (9.1)        \\
SSD-VGG                          & 0.11 (16.1)  & 0.23  (16.5)     & 0.33 (25.7)    & 0.51 (44.6)        \\
\bottomrule
\end{tabular}%
\vspace{-6pt}
\caption{Memory (GB) and time (ms) requirements for loading/running inference with 3 different batch sizes (in frames). Run memory values include load values, but exclude memory needs of serving frameworks. Results use a Tesla P100 GPU.}
\label{tab:models_memory_and_time_requirements}
\end{table}

Table~\ref{tab:models_memory_and_time_requirements} breaks this memory pressure down further by listing the amount of loading and running memory required for representative models in our workloads. When analyzed in the context of the scale of edge video analytics workloads, the picture is bleak, even with a batch size of 1. For example, a 2 GB edge box can support only 1, 2, or 3 VGG16, YOLOv3, or ResNet50 models, respectively, after accounting for the memory needs of the serving framework. Moving up to 8 and 16 GB edge boxes (of course) helps, but not enough, \eg an 8 GB box can support 13 YOLOv3 or 2 Faster RCNN models, both of which are a drastic drop from the 10s of models that workloads already involve (\S\ref{s:workloads}).

\subsection{\fontsize{10.5}{10}{Limitations of Existing GPU Memory Management}}
\label{ss:limitations}
\vspace{-4pt}

\para{Space and time sharing.} Existing learning frameworks recommend model allocation at the granularity of an entire GPU~\cite{DNNWorkloadAnalysisATC19}. Space-sharing techniques~\cite{MPS,TensorRT} eschew this exclusivity and partition GPU memory per model. Although space-sharing approaches are effective when a workload's models can fit together in GPU memory, they are insufficient when that does not hold, which is common at the edge (\cref{ss:problem}) 

There are two natural solutions when a workload's models cannot fit together in the target GPU's memory. The first is to place models on \emph{different} GPUs~\cite{nexus, clockwork}, which resource-constrained edge settings cannot afford. The second is to \emph{time share} the models' execution in the GPU by \emph{swapping} them in and out of GPU memory (from CPU, via a PCIe interface)~\cite{SwapAdvisor, nexus, antman, pipeswitch, clockwork}. 
However, as we will show next, time-sharing techniques are bottlenecked by frequent model swapping, which severely limits their utility. More recently, SwapAdvisor~\cite{SwapAdvisor} and Antman~\cite{antman} proposed swapping at finer granularities, e.g., individual or a few layers. However, even these approaches are limited in our case because a handful of layers in vision DNNs typically account for most memory usage (\S\ref{ss:observations}); edge boxes often lack the GPU memory to concurrently house even these expensive singular layers.

We evaluated time-sharing strategies in our setting by considering a hybrid version that \emph{packs} models into GPU memory, and executes as many models as possible while ensuring that swapping costs for the next model to run are hidden. Concretely, we extend Nexus~\cite{nexus} to incorporate such pipelining. Our variant first organizes models in round-robin order (as Nexus does), and profiles the workload offline to determine the best global list of per-model batch sizes that maximizes the minimum achieved per-model throughput while adhering to an SLA (i.e., a per-frame processing deadline). Using those batch sizes, the scheduler traverses the round robin order with the goal of minimizing GPU idle time: when loading the next model, if there does not exist sufficient memory to load both parameters and intermediates, the most recently run model (i.e., the one whose next use is in the most distant future) is evicted to make space.

\begin{figure}[!tbp]
\center
    \includegraphics[width=0.79\linewidth]{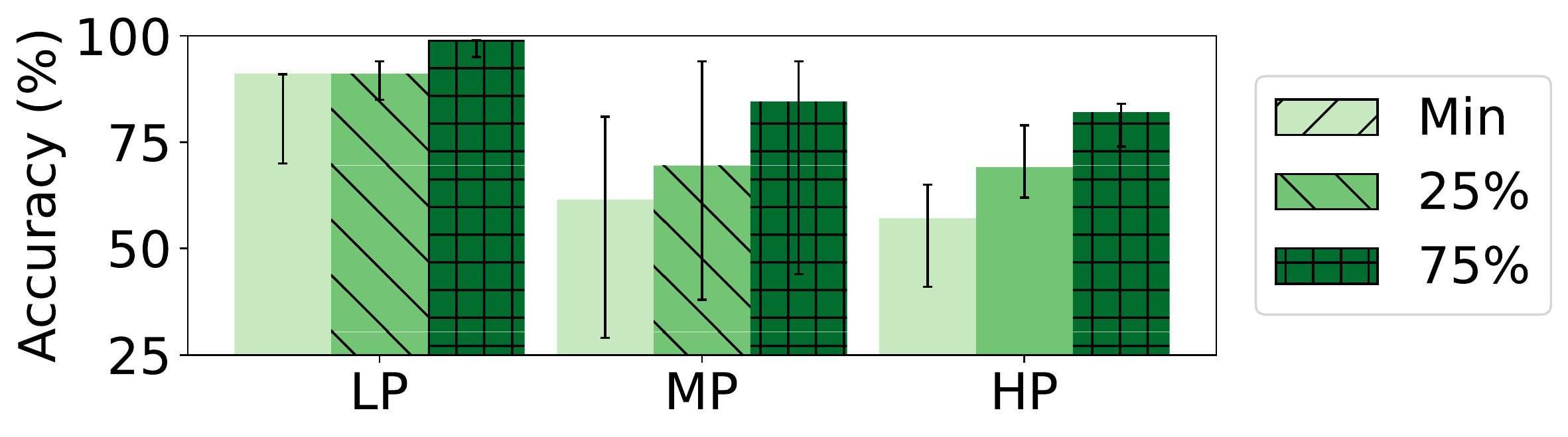}
    \vspace{-10pt}
    \caption{Achieved accuracy with time/space-sharing alone (\ie using our Nexus variant) for different memory availability (following the definitions in \S\ref{s:workloads}). Bars list results for the median workload in each class, with error bars spanning min to max.}\vspace{-4pt}
    \label{fig:nexus}
  \end{figure}

Figure~\ref{fig:nexus} shows the accuracy of the Nexus variant on our workloads with an SLA of 100 ms; we saw similar trends for other common SLAs in video analytics~\cite{nexus}. As shown, accuracy drops are substantial, growing up to 43\% relative to a setting when there exists sufficient memory to house all models at once. 
The root cause is the disproportionately high loading times of vision DNNs that must be incurred when swapping. As shown in Table~\ref{tab:models_memory_and_time_requirements}, per-model loading delays are 0.98-34.4$\times$ larger than the corresponding inference times (for batch size 1). When facing the strict SLAs of video analytics, these loading costs result in the inability to keep pace with incoming frame rates, and thus, dropped (unprocessed) frames; the Nexus variant skipped \fillin{19-84}\% of frames. 

\para{Predicting workload characteristics.} Another approach is to selectively preload models based on predictions of the target workload~\cite{distream}, \eg deprioritizing inference on streams at night due to lack of activity. However, in edge video analytics, spatial correlation between streams results in model demands being highly correlated~\cite{spatula,caesar,reducto,chameleon}.

\para{Compression and quantization.} These techniques generate lighter model variants that impose lower memory (and compute) 
footprints and deliver lower inference times. Some families offer off-the-shelf compressed variants (\eg Tiny YOLOv3), and techniques such as neural architecture search can be used to develop cheaper variants that are amenable to deployment constraints~\cite{mistify}. Regardless, in reducing model complexity, these cheaper model variants typically sacrifice accuracy and are more susceptible to drift, relative to their more heavy-weight counterparts~\cite{ekya,odin}; consequently, determining the feasibility of using such models in a given setting requires careful tuning and analysis by domain experts.

We consider compression and quantization as orthogonal to merging for two reasons. First, in common workloads that involve a mix of models and tasks (\S\ref{s:workloads}), it may not be possible to compress all of the models while delivering sufficient accuracy. However, even a handful of non-compressed models can exhaust the available GPU memory (\S\ref{ss:problem}). Second, compressed models exhibit sharing opportunities: our workloads include compressed and non-compressed models (\S\ref{s:workloads}), and our results show that \name{} is effective for both (\S\ref{sec:eval}).

\section{Our Approach: Model Merging}
\label{s:merging}

To address the high model loading costs that plague existing memory management strategies when workloads cannot fit together in a GPU's memory (\S\ref{ss:limitations}), we propose \emph{model merging}. Merging is complementary to time/space sharing of GPU memory, and its goal is straightforward: share layers across models such that only one copy of each shared layer (i.e., layer definition and weights) must be loaded into GPU memory and can be used during inference for all of the models that include it. The benefits are two-fold: (1) reduce the overall memory footprint of a workload, thereby enabling edge boxes to house more models in parallel and perform fewer swaps (or equivalently, lower the number of edge boxes needed to run the workload), and (2) accelerate any remaining swaps by reducing the amount of extra memory that the next model to load requires. Note that merging does not involve sharing intermediates (\ie layer outputs) for a common layer because models may run on different videos (and thus, inputs). We next highlight the promise for merging in edge video analytics (\S\ref{ss:potential}), and then lay out the challenges associated with realizing merging in practice (\S\ref{ss:challenges}).

\subsection{Opportunities}
\label{ss:potential}

 \vspace{-5pt}
\para{Commonality of layers.} \add{A layer is characterized by both its architecture and its weights. 
In ML frameworks (\eg PyTorch, TensorFlow), the architecture is defined by first specifying a layer type (\eg convolutional, linear, batch normalization), which in turn indicates how the layer transforms inputs, and dictates the set of defining parameters that must be specified (\eg convolutional: kernel, stride, etc., linear: \# of input features, bias, etc.). A layer's weights are a matrix of numbers whose dimensions match the layer structure. 
To successfully share a layer across a set of models, that layer must be \emph{architecturally} identical in each model, but its weights need not be the same across appearances.}

\begin{figure}[t]
  \centering
  \includegraphics[width=.85\linewidth]{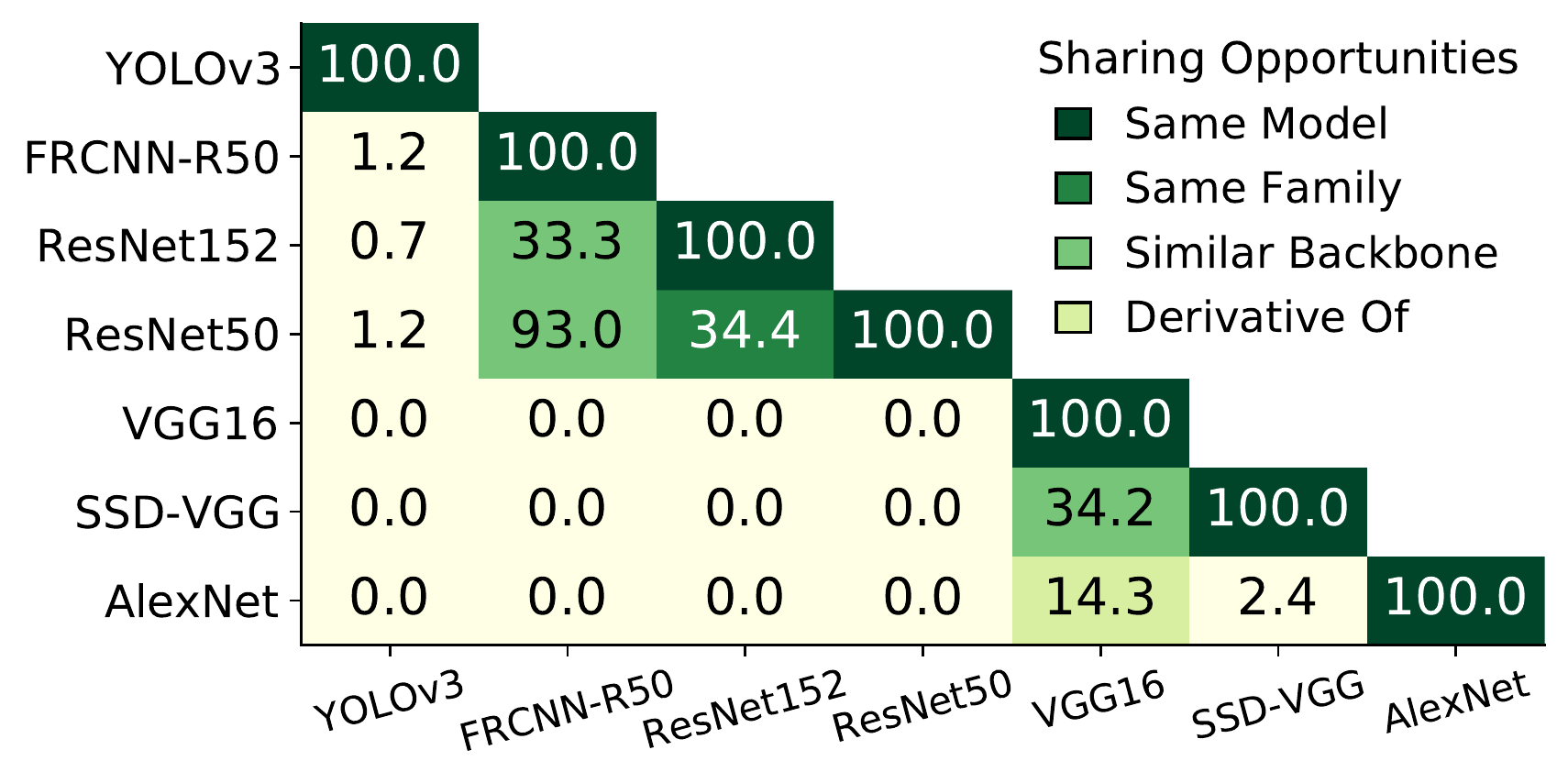}
  \vspace{-10pt}
  \caption{\add{Percentage of architecturally identical layers across different model pairs. See Figure~\ref{fig:model_overlap_complete} for an extended version.}}
  \label{fig:models_overlap}
\end{figure}

\add{Architectural equivalence is determined directly from the model definition in the ML framework (i.e., no inference required): the layers must be of the same type, with identical values for type-specific properties. Using this approach, we studied pairs of 24 different models to identify and analyze layers with identical architectures. \S\ref{ss:additional_figures} and Figures~\ref{fig:deep_dive_sharing1}-\ref{fig:deep_dive_sharing3} present our comprehensive results and break down sharing opportunities by layer type. Below, we summarize our findings; Figure~\ref{fig:models_overlap} lists results for representative model pairs.} 

Model pairs fall into one of three categories: (1) instances of the same model, (2) different models in the same family (\eg ResNet \add{variants}), and (3) different models in different families. Multiple instances of the same model unsurprisingly match on every layer; this favorable scenario is not uncommon in edge video analytics, as several model architectures tend to dominate the landscape~\cite{TrafficCaseStudy} and a given model can be employed on different video feeds or in search of different objects (\S\ref{s:workloads}). \add{More interestingly, we also observe sharing opportunities across different models from the same (up to \fillin{84.6\%}) and divergent (up to \fillin{96.3}\%) families.}

\add{Models within the same family exhibit significant sharing opportunities as larger variants are typically extended versions of the original base model. For instance, ResNet models share ResNet blocks (groups of 2-3 convolutional layers) that are repeated at different frequencies, as well as the first convolutional layer and final fully-connected layer. As a result, all 41 layers of ResNet18 are shared with ResNet34 (Figure~\ref{fig:deep_dive_sharing3}). Similarly, in the VGG family, models share the exact same base architecture and add different numbers of convolutional layers, e.g., VGG19 shares all 16 of VGG16's layers (13 convolutional and 3 fully-connected; Figure~\ref{fig:deep_dive_sharing1}).}

\add{Sharing for models in different families comes in two main forms: (a) `similar backbones' and (b) `derivatives of.' Scenario (a) includes pairs of detectors that use the same (or similar) backbone networks for feature extraction, e.g., SSDs that use any VGG backbone, or FasterRCNNs that use any ResNet backbone. (a) also includes pairs of classifiers and detectors where the classifier (or a close variant) is used as the detector's backbone. For instance, every layer in the ResNet50 backbone of FasterRCNN (which constitutes 51\% of the detector's layers) appears in the ResNet101 classifier. Similar examples include SSD-VGG with any VGG variant, and SSD-MobileNet with MobileNet. Scenario (b) involves cases where one model family was derived directly from another. For example, VGG was developed by replacing AlexNet's large kernels with multiple smaller ones~\cite{vgg}; VGG16 and AlexNet share 3 out of 16 layers, including 2 fully-connected layers at the end (Figure~\ref{fig:deep_dive_sharing2}). Other examples include InceptionNetV3~\cite{inceptionnetv3} with GoogLeNet~\cite{googlenet}.}

\add{In summary, \fillin{43}\% of all pairs of different models present sharing opportunities. Of those with substantial  ($\geq$ 10\%) common layers, 51\% have models in the same family, while 49\% involve models from different families; for the latter, 76\% are `similar backbones' and 24\% are `derivatives of.'}

These layer similarities \add{generally} follow from the fact that the considered models are all vision processing DNNs. That is, they all ingest pixel representations of raw images, and employ a series of traditional CV operations~\cite{visionsurvey1,visionsurvey2}, \eg a convolutional layer is applying a learned filter to raw pixel values in preparation for downstream processing. Moreover, the target tasks are rooted in identifying and characterizing objects in the scene using low-level CV features such as detected edges and corners~\cite{background-subtraction-cvpr2011,event-camera-eccv16,inertial-odometry-cvpr17,feature-tracks-iros16,probabilistic-data-icra17, reducto,FocusOSDI2018}. 

Prior work has capitalized on such similarities for efficient multi-task learning~\cite{caruana1997multitask,multi1,MainstreamATC2018} and architecture search~\cite{darts,arch2}. Those efforts aim to reduce computation overheads by sharing ``stems'' of models, \ie contiguous layers (and their generated intermediates) starting from the beginning of the models. In contrast, we aim to exploit architectural similarities to reduce memory overheads via merging. As a result, merging only requires layer definitions and weights to be shared, but not generated intermediate values. This distinction is paramount because, as we will discuss in \S\ref{ss:observations}, memory-heavy layers typically reside towards the end of vision DNNs. Consequently, stem sharing would require almost all model layers to be shared to reap substantial memory savings, which in turn brings unacceptable accuracy drops (\S\ref{ss:challenges} and \S\ref{sec:eval}). 
Merging, on the other hand, can share only those memory-heavy layers to simultaneously deliver substantial memory savings and preserve result accuracy.

\begin{figure}[!tbp]
  \centering
  \includegraphics[width=0.82\linewidth]{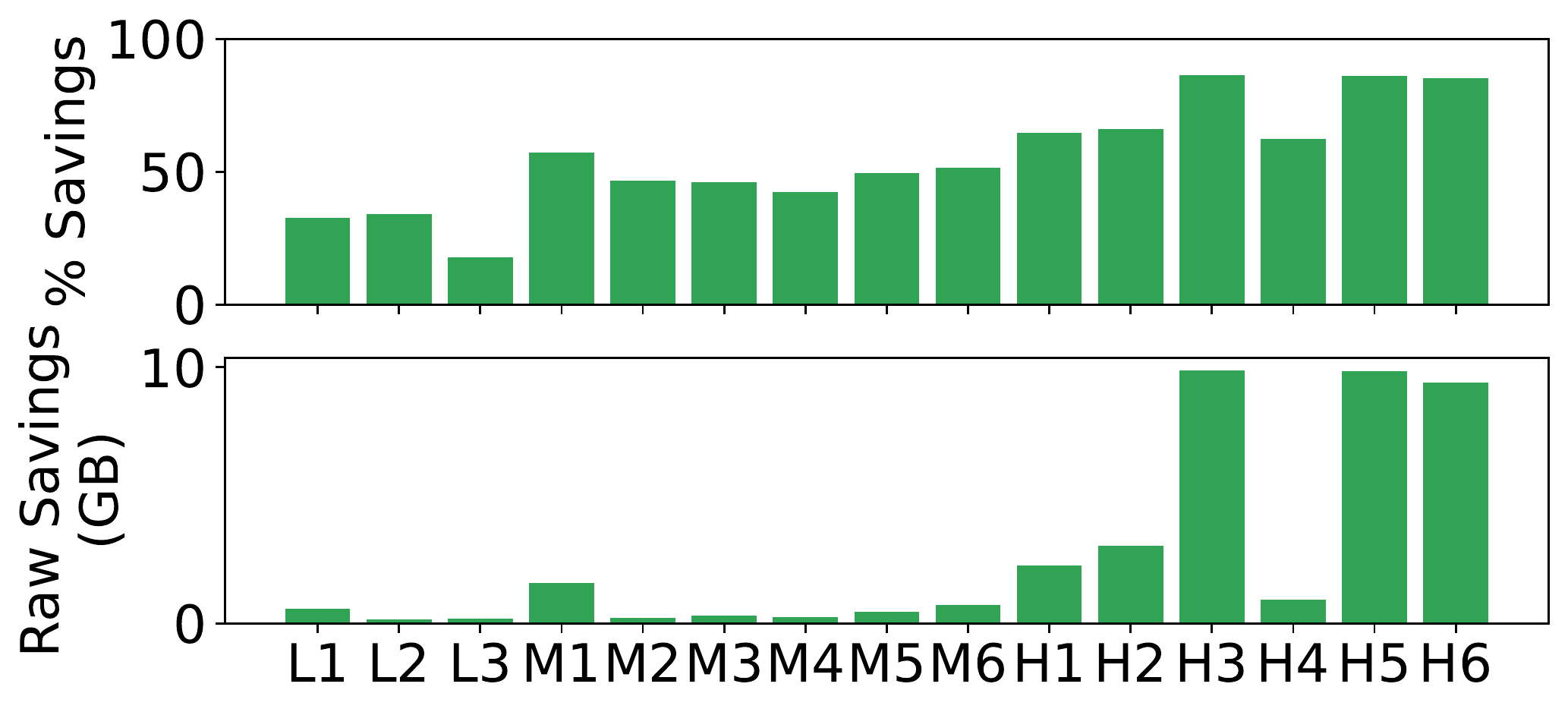}
  \vspace{-10pt}
  \caption{Potential memory savings when all architecturally identical layers are shared across the models in each workload.}
  \vspace{-7pt}
  \label{fig:upper_memory}
\end{figure}

\begin{figure}[!tbp]
  \centering
  \includegraphics[width=0.77\linewidth]{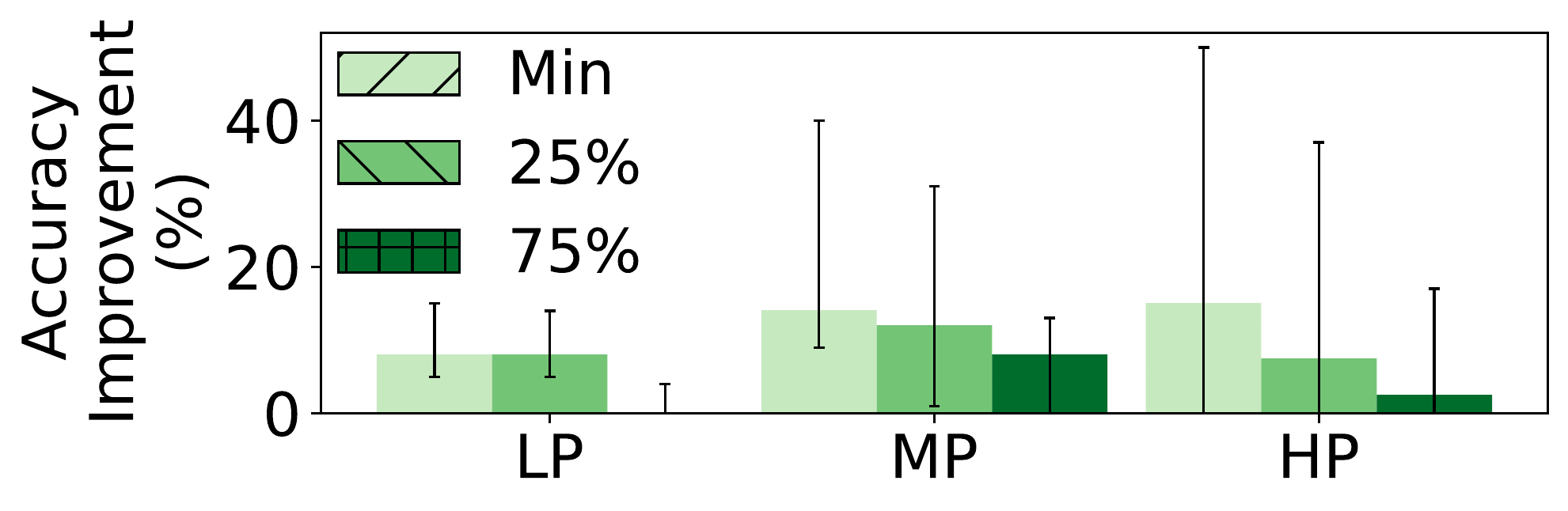}
  \vspace{-10pt}
  \caption{Potential accuracy improvements when sharing all architecturally identical layers. Memory availability is defined in \S\ref{s:workloads}, bars list medians, and error bars span min to max.}
  \vspace{-1pt}
  \label{fig:upper_acc}
\end{figure}

\para{Potential memory savings and accuracy improvements.}
Figure~\ref{fig:upper_memory} shows the memory savings from sharing all of the common layers across the models in each of our workloads; this represents an \emph{upper bound} on merging benefits as it disregards the challenge of identifying an acceptable set of weights per shared layer (\S\ref{ss:challenges}). As shown, the memory savings are substantial: per-workload memory usage dropped by \fillin{17.9-86.4}\% relative to no merging, translating to raw savings of \fillin{0.2-9.9} GB. Importantly, these savings result in \fillin{2} and \fillin{4} new workloads fitting entirely (no swapping) on edge boxes with 2 GB and 8 GB of GPU memory (with batch size 1). Similarly, the number of 2 GB edge boxes needed to support each workload drops from \fillin{1-9 to 1-4}. We further evaluated the resulting impact on end-to-end accuracy by comparing the performance of the Nexus variant from \S\ref{ss:limitations} when run on workloads with and without (maximal) merging. Models in both cases were ordered in the same way, to maximize the benefits of merging (\S\ref{ss:edge}). As shown in Figure~\ref{fig:upper_acc},
merging can boost accuracy by up to 50\% across our workloads. 
These benefits are a direct result of lower swapping costs, and the resulting ability to run on \fillin{29-61}\% more frames. 

\subsection{Challenges}
\label{ss:challenges}

Merging layers for memory reductions requires using shared weights across the models in which those layers appear. However, those shared weights must not result in accuracy violations for any of the models, despite their potentially different architectures/tasks, target objects/videos, etc.; such accuracy drops would forego merging benefits from faster swapping.
Concretely, there are two core challenges in practically exploiting the architectural commonalities from \S\ref{ss:potential}.

\begin{figure}[!tbp]
  \centering
  \includegraphics[width=0.7\linewidth]{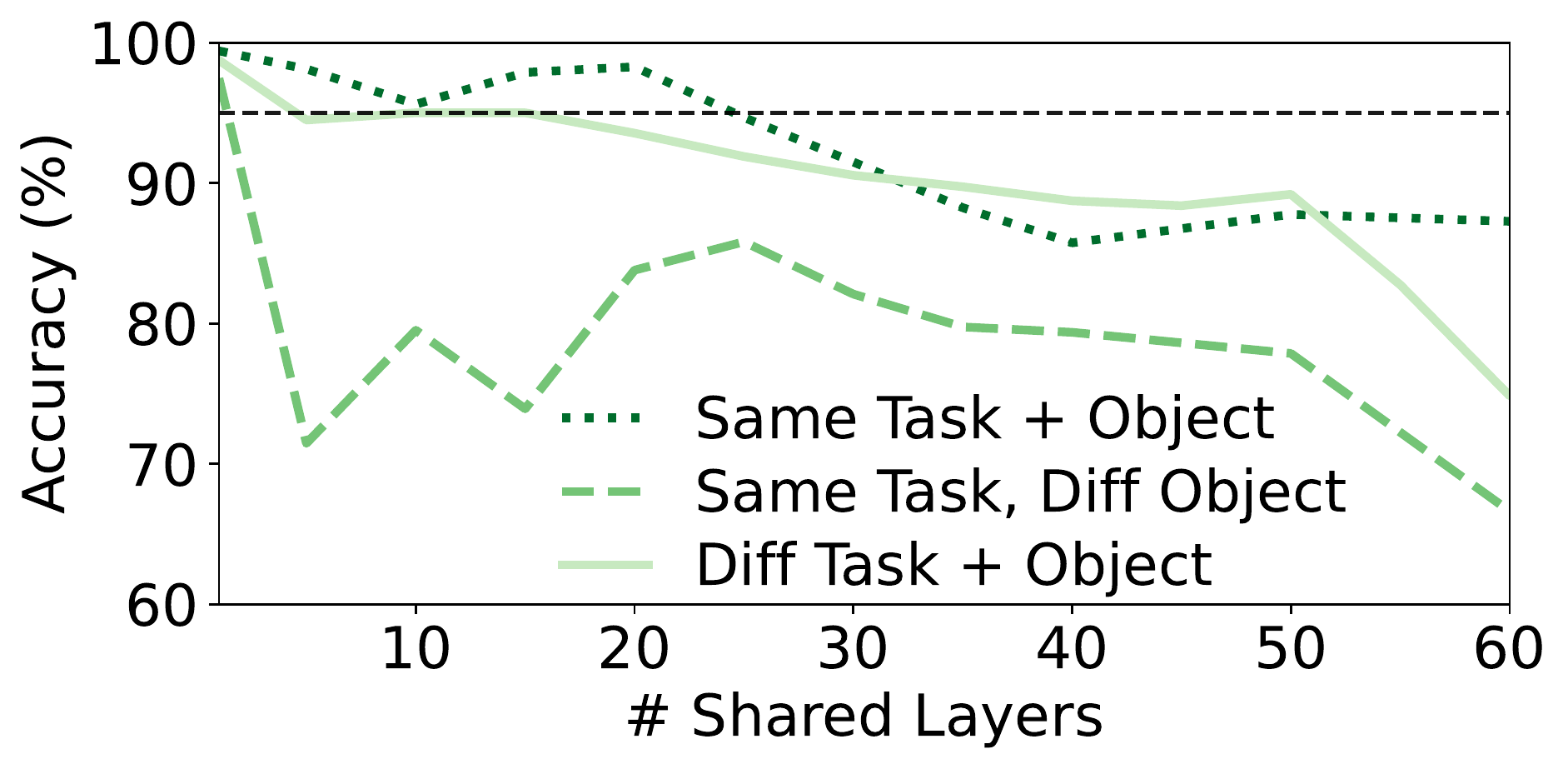}
  \vspace{-10pt}
  \caption{Accuracy after 5 hours of retraining when sharing additional architecturally-identical layers for different model pairs  (starting from their origins). Tasks cover detection (Faster RCNN) and classification (ResNet50), and two objects: people, vehicles. Results list the lower per-model accuracies per pair.} 
  \label{fig:sharing_acc}
\end{figure}

\para{Challenge 1: sharing vs. accuracy tension.} To maximize memory savings, merging seeks to share as many architecturally identical layers as possible across a workload's models. However, we observe that accuracy degradations steadily grow as the number of shared layers increases. Figure~\ref{fig:sharing_acc} illustrates this trend by sharing different numbers of identical layers across representative pairs of models that vary on the aforementioned properties, \eg target task. These results were obtained when we increase the number of shared layers by moving from start to end in the considered models, but  similar trends are observed for other selection strategies (\eg random) and models. 

The reason for this behavior is intuitive: the retraining performed to assess the feasibility of a sharing configuration is \emph{end-to-end} across the considered models. During this process, weights are being tuned for all of the layers in all of the models, with the constraint being that the shared layers each use a unified set of weights. Sharing more layers has three effects: (1) more constraints are being placed on the training, (2) it is harder to find weights for (the shrinking number of) unshared layers that simultaneously accommodate the growing constraints, and (3) learning each model's desired function becomes more difficult as there exist fewer overall parameters to tune~\cite{overparam1,overparam2}. It is for these reasons that isolated merging strategies such as averaging weights across copies of each shared layer (while keeping other layers unchanged) do not suffice; we find that sharing even single layers in this way almost always results in unacceptable accuracy dips.

Digging deeper, the issue stems from non-convex optimization of DNNs, which leads to several equally good global minima~\cite{nonlin1,nonlin2}. Thus, training even two identical models on the same dataset, and for the same task/object, often results in divergent weights across each layer, despite the resultant models exhibiting similar overall functionality. 

\para{Challenge 2: retraining costs.} The retraining involved in determining whether a set of layers to share can meet an accuracy target, and if so, the weights to use, can be prohibitively expensive. For instance, each epoch when jointly retraining two Faster RCNN models that detect cars at nearby intersections (\ie a simple scenario) took $\approx$35 mins, and different combinations of layer sharing required between 1-10 epochs to converge. These delays grow as more models are considered since training data must reflect the behavior of all of the unmerged models that are involved, \eg by using the original training datasets for each of those models. Worse, it is difficult to know, a priori, which sharing configurations can meet accuracy targets (and which will not) in a reasonable time frame. For example, the model pairs in Figure~\ref{fig:sharing_acc} have largely different `breaking points.' 
The result also fails to support the use of intuitive trends to predict the success of sharing configurations: models targeting the same task or object do not exhibit any discernible advantage.
\section{\name Design}
\label{s:gemel}

\name{} is an end-to-end system that practically integrates model merging into edge video analytics pipelines by addressing the challenges in \S\ref{ss:challenges}. We first provide an overview of \name{}'s operation, and then describe the core observations (and resulting optimizations) that it leverages to enable timely merging without violating accuracy requirements.

\subsection{Overview}
\label{ss:overview}
\begin{figure}[!tbp]
  \centering
  \includegraphics[width=0.9\linewidth]{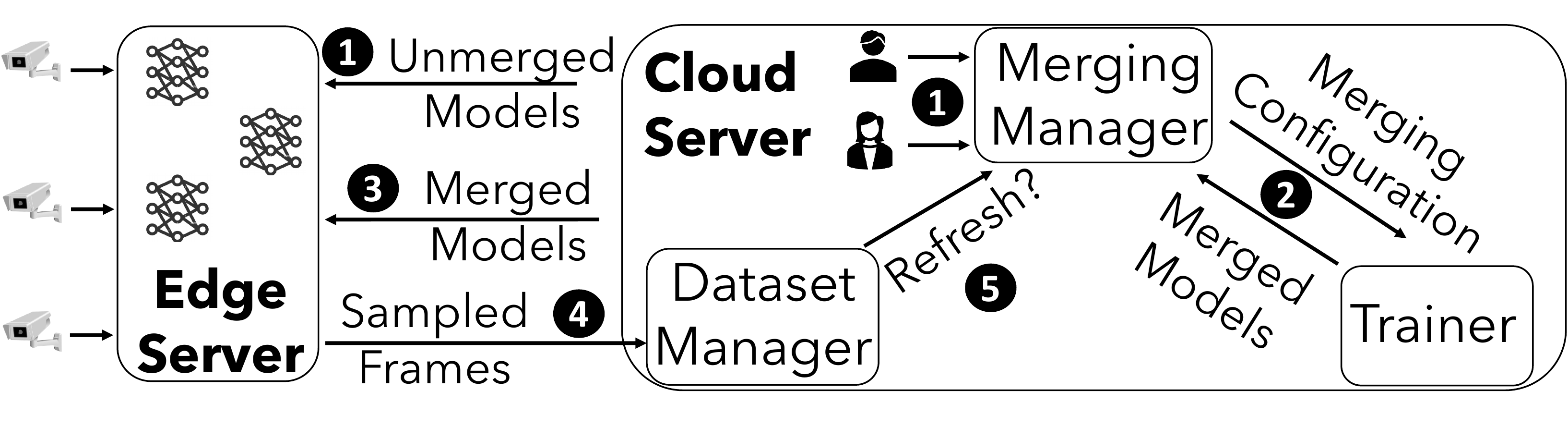}
\vspace{-10pt}
  \caption{\name architecture.}
  \vspace{-4pt}
  \label{fig:architecture}
\end{figure}

Figure~\ref{fig:architecture} shows \name{}'s cloud merging and edge inference workflows. 
As in existing pipelines~\cite{rocket,chameleon,reducto}, users register inference tasks (or ``queries'') at \name{}'s cloud component by providing a DNN, and specifying the input video feed(s) to run on as well as the required accuracy for the results.
Upon receiving new queries, \name{} bootstraps edge inference by sending unaltered versions of the registered models to the appropriate edge box(es) \numcircledtikz{1}. When GPU memory is insufficient to house all of those models, edge boxes run the Nexus variant from \S\ref{ss:limitations} that pipelines inference and model loading to maximize the min per-model throughput.

After initiating edge inference, \name{}'s cloud component begins the merging process, during which it \emph{incrementally} searches through the space of potential merging configurations across the registered models, and evaluates the efficacy of each configuration in terms of both its potential memory savings and its ability to meet accuracy requirements \numcircledtikz{2}. The evaluation of each configuration involves joint retraining and validation of the models participating in merging. Since \name{}'s goal is to ensure that the retrained models deliver sufficient accuracy (relative to the originals) on the target feeds, data for these tasks can be obtained in one of two ways: users can supply the data used to train the original models, or \name{} can automatically generate a dataset by running the supplied model (or a high-fidelity one~\cite{chameleon,VideoStormNSDI2017}) on sampled frames from the target feed.

At the end of each merging iteration, if the considered configuration was successfully retrained to meet the accuracy targets for all constituent models, \name{} shares the updated merged models with the appropriate edge boxes \numcircledtikz{3}. New merging results may result in altered edge inference schedules to maximize merging benefits for reducing swapping costs and boosting inference throughput. The iterative merging process for the current workload then continues until (1) the cloud resources dedicated to merging have been expended, (2) no configurations that can deliver superior memory savings are left to explore, or (3) models with sharing opportunities are either newly registered or deleted by users.

\name{} periodically assesses \emph{data drift} for its merged models. As in prior systems~\cite{odin,reducto}, edge servers periodically send sampled frames (and their inference results, if collected) to \name{}'s cloud component \numcircledtikz{4}. These sampled frames are used to augment the datasets considered for retraining merged models, and to track the accuracy of recent results generated at the edge by deployed merged models. For the latter, \name{} runs the original user models on the sampled videos and compares the results to those from the merged models. If accuracy is below the target for any query, \name{} reverts edge inference to use the corresponding original (unmerged) models, and resumes merging and retraining, starting with the previously deployed weights \numcircledtikz{5}.

\para{Implementation.} \name{} uses PyTorch~\cite{PyTorch} to manage cloud merging and edge inference, and is implemented in $\approx$3500 LOC. \add{More details are presented in \ref{ss:implementation_details}.}

\subsection{Guiding Observations}
\label{ss:observations}

Two key empirical observations guide \name{}'s approach to tackling the challenges in \S\ref{ss:challenges}. We describe them in turn.

\begin{figure}[!tbp]
  \centering
  \includegraphics[width=0.9\linewidth]{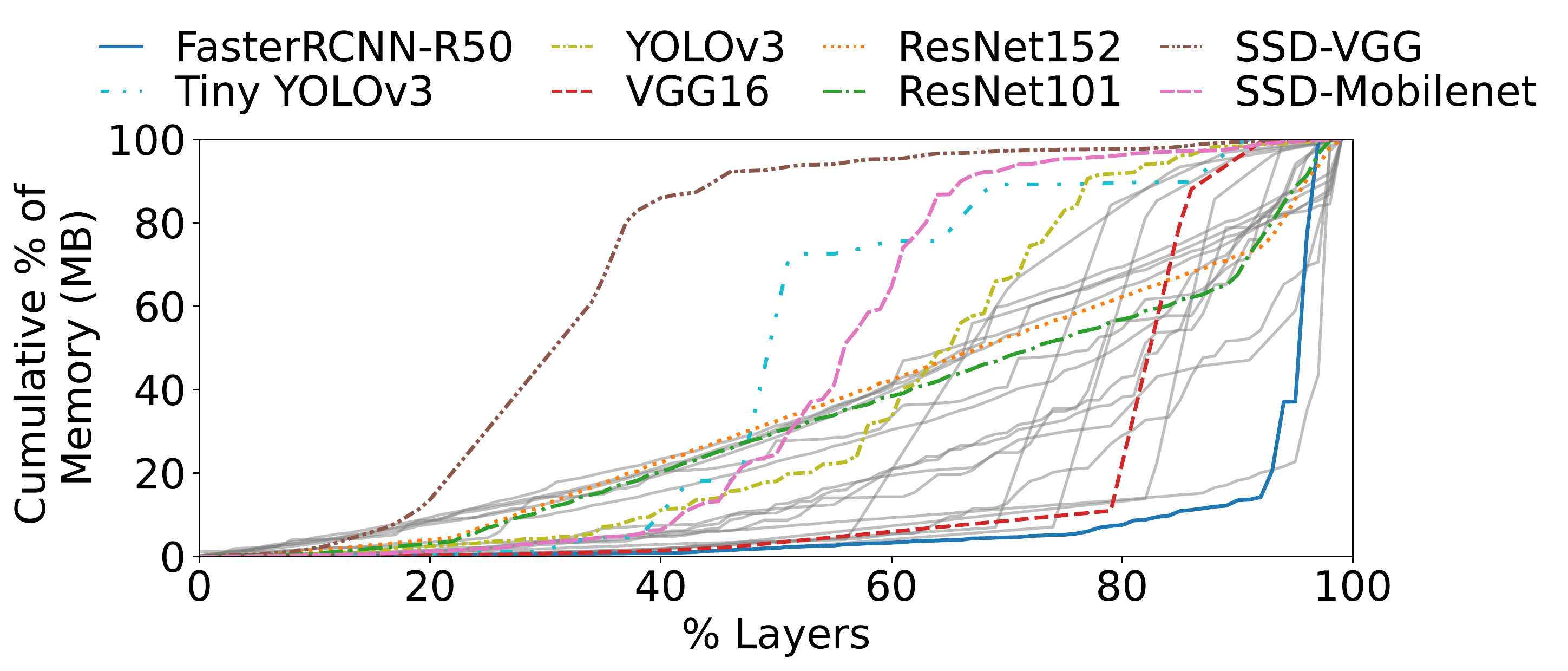}
  \vspace{-10pt}
  \caption{\add{Cumulative memory consumed by each model's layers moving from start to end of the model. \S\ref{ss:additional_figures} has full legend.}}
  \label{fig:mem_by_layer_maintext}
\end{figure}

\para{Observation 1: power-law memory distributions.} We find that vision DNNs \add{commonly} exhibit power-law distributions in terms of memory usage, whereby a few ``heavy-hitter'' layers account for most of the overall model's memory consumption. \add{Figure~\ref{fig:mem_by_layer_maintext} illustrates this, showing that for 80\% of considered models, 15\% of the layers account for 60-91\% of memory usage. For example, a single layer in VGG16 is responsible for 392 MB (the entire model is 536 MB) and corresponds to the steep slope around the x=80\% mark. Similarly, Tiny YOLOv3 has three layers (around the 38\%, 45\%, and 65\% marks) that together use 35 MB of its total 42 MB.}

\add{Heavy-hitter layers come in one of two forms. The first are the convolutional layers at the end of the feature extractor that condense the numerous low-level features extracted by prior layers (e.g., shapes, colors) into higher-level, more abstract features (e.g., eyes, nose). The second are the subsequent fully-connected layer(s) that learn more robust patterns from all possible combinations of those high-level features, e.g., eyes, nose, and fur might each suggest a dog, but the combination is a stronger indicator. Note that models generally include one such fully-connected layer per sub-task, e.g., detectors have one for finding bounding boxes and one for classifying objects. Memory-heavy fully-connected layers are spatially close to one another (within a few layers), and are usually followed by 1-2 cheap fully-connected layers that extract predictions from the final feature vector.}

\add{The main exception is ResNet, whose models use residual layers to address accuracy saturation limitations of prior deep models~\cite{resnet}. ResNet models have memory-heavy ResNet blocks (set of convolutional layers) that repeat at varying frequencies, thereby distributing memory more evenly across the models, e.g., ResNet101 and ResNet152 repeat the same ResNet block 23 and 36$\times$, leading to gradual slopes in Figure~\ref{fig:mem_by_layer_maintext}.
DenseNet has the same pattern~\cite{densenet}.}

\add{Figure~\ref{fig:mem_by_layer_maintext} also shows that heavy-hitter layers most often appear in the latter half of a model's architecture (since both forms involve condensing features from earlier layers), complicating the use of stem sharing for memory savings (\S\ref{ss:challenges}). 
For example, Faster RCNN’s expensive fully-connected layers fall at layers 101 and 104 out of 106, and together account for 76\% of total memory. The few cases with heavy-hitters in the middle of a model (between the 20-60\% marks) are ``single-shot'' detectors (SSD-VGG, SSD-Mobilenet, Tiny YOLOv3, YOLOv3) that find bounding boxes and classify objects at once, rather than as disparate subtasks. These models replace the few memory-heavy fully-connected layers (for those subtasks) with many cheaper convolutional layers; doing so extends model lengths and shifts the large jump from memory-heavy \emph{feature extractor} layers to earlier.}

\add{These observations have two implications on merging. First, strategies can reap most potential memory benefits by targeting the few heavy-hitter layers in models. Thus, the tension between memory savings and accuracy is far more favorable than that between the number of shared layers and accuracy (Figure~\ref{fig:sharing_acc}). Second, strategies should be agnostic to the position of heavy hitters in models, and must support the common case where heavy hitters appear towards the end.}

\para{Observation 2: independence of per-layer merging decisions.} In DNNs, layers are configured based on input formats, target task, execution time, etc.
Hence, a natural assumption is that the ability to share any one layer is dependent on sharing decisions for other layers, \eg a layer may be shareable if and only if other layers are shared. Prior work has highlighted that inter-layer dependencies primarily arise between neighboring layers, \eg with transfer learning, performance drops are largest when splitting neighboring layers~\cite{multi1}. \add{Thus, to determine the existence of layer-wise dependencies as it pertains to merging, we focus our analysis on (potential) dependencies between neighboring layers; we also consider other layers via random selection. Using the 25\% most memory-heavy layers for each model in our workloads, we test whether accuracy targets are met under different sharing configurations (described in Table~\ref{t:independence_tab}).} 

\add{As shown, we \emph{never} observe a case where a layer is unable to meet an accuracy target on its own, but it is able to meet the accuracy target when some other layers are also shared (shaded row in Table~\ref{t:independence_tab}). This is consistent with our finding that sharing more layers leads to larger accuracy degradations (Figure~\ref{fig:sharing_acc}) since additional constraints are placed on the weights for those layers, and fewer (unconstrained) non-shared layers exist to help satisfy the constraints. The implication is that layers can be considered independently during merging without harming their potential merging success.}

\newcolumntype{a}{>{\columncolor{YellowGreen}}p}
\begin{table}[tp!]
\scriptsize
\centering
\begin{tabular}{|p{1.4cm}|p{1.3cm}|a{1.7cm}|p{0.5cm}|p{0.75cm}|}
\hline
& \textbf{Only Alone} & \textbf{Only Alternate} & \textbf{Both} & \textbf{Neither}\\
\hline
\textbf{1 Each Side} & 1.1\% & 0.0\% & 97.6\% & 1.3\% \\
\hline
\textbf{2 Each Side} & 3.7\% & 0.0\% & 95.0\% & 1.3\% \\
\hline
\textbf{Random} & 8.5\% & 0.0\% & 90.2\% & 1.3\% \\
\hline
\end{tabular}
\vspace{-6pt}
\caption{\add{Sharing a layer alone vs. \emph{alternate} approaches (sharing a layer with one or two neighbors on each side, or with 3 random sets of 1-10 layers). Results are \% of runs that meet accuracy targets (aggregated across 80, 90, 95\%), and list cases where the layer alone met but an alternate did not, an alternate met but the layer alone did not, both met, and neither met.}}
\label{t:independence_tab}
\end{table}

\para{Takeaway.} Collectively, these observations motivate an incremental merging process (detailed in \S\ref{ss:heuristic}) that attempts to share one new layer at a time, and prioritizes heavy-hitter layers that consume the most memory (and are thus the most fruitful to share). In this manner, memory-heavy layers are considered in the most favorable settings (\ie with the fewest other shared layers), and each increment only modestly adds to the likelihood of not meeting accuracy targets.

\para{Note.} Despite arising across our diverse workloads, these observations are not guarantees. Importantly, violation of these observations only results in merging delays (inefficiencies), but not accuracy breaches; accuracy is explicitly vetted prior to shipping merged models to the edge for inference.

\subsection{Merging Heuristic}
\label{ss:heuristic}
\name{} begins by enumerating the layers that appear in a workload, and annotating each with a listing of which models the layer appears in (and where) and the total memory it consumes across the workload; we refer to all appearances of a given layer as a `group.' \name{} then sorts this list in descending order of memory consumption, \eg a 100 MB layer that appears in 4 models would be earlier than a 120 MB layer that appears 3 times. Thus, memory-heavy groups, or those that would yield the largest memory savings if successfully merged, are towards the start of the list. 

\name{} then maintains a running merging configuration, and simultaneously merges and trains layers across models in an \emph{incremental} fashion. To begin, \name{} selects the first group from the sorted list (\ie the one that consumes the most memory in the workload) and attempts to share it across all of the models in which it appears; this group is added to the running configuration. While a subset of models could be considered instead, \name{} aggressively opts to first try sharing across all models in the group, and then to selectively remove appearances of the layer when the resulting accuracy is insufficient. The reason is that we did not observe any model clustering strategies (\eg based on task) that identified models consistently unable to share layers.

To retrain and merge the current running configuration, \name{} selects initial weights for the newly added group from a random model that includes that layer. \add{We tried selecting weights from each model (including the one with the highest accuracy) but found no difference in the \# of epochs needed to meet accuracy.  
We also tried default initialization techniques (e.g., Kaiming\cite{kaiminginitialization}), which led to lower accuracy.} Retraining continues until the merged models each meet their accuracy targets, or a preset time budget elapses (10 epochs by default). If retraining is successful, \name{} adds the next group in the sorted list to the running configuration, and resumes retraining from the weights at the end of the previous iteration. The generated merged models are sent to the edge box and incorporated into edge inference (\S\ref{ss:edge}).

If retraining is not successful at the end of an iteration, \name{} must decide whether to prune \add{layers} from the current group and try again, or to discard the group altogether and move on to the next one in the sorted list. \add{To do this, \name{} follows a strategy that aims to balance fast memory savings and avoidance of unsuccessful training rounds, with priority on the latter since failures can consume 3-10 epochs (each up to 30 min) and provide no new memory savings. Specifically, recall that each time a new group is considered, the number of shared layers in the merging configuration grows by the size of the group. To counter this `additive increase,' upon unsuccessful retraining, \name{} halves the current group, eliminating half of the layer appearances. If the resulting layer appearances consume more memory than the next group, \name{} considers those layers; else, \name{} removes the current group from the running policy, and moves to the next one. In either case, retraining resumes from the weights at the end of the last successful iteration.
We compare against alternate merging heuristics in \S\ref{ss:deepdive}.}

\para{Accelerating retraining.}
Each iteration requires \name{} to run retraining over many epochs, and validate the results accuracy-wise. 
To accelerate training and validation, \name{} takes an adaptive approach. During validation, as per-model accuracy values approach their targets, it is
often unnecessary to train further on full epochs of data. Instead, \name{} reduces
the training data once the accuracy is within a pre-defined threshold
of the target. Specifically, \name{} reduces the amount of data so it is inversely proportional to the gap in accuracy normalized by the lift since the previous training. Reducing data on such \emph{early success} directly translates to
lower training times. Similarly, \name{} detects \emph{early failures} by
looking at the validation results and removing models that are not improving at the same pace as the others after some time (3 epochs by default). 
We empirically observe that early success and early failure detection drastically (28\% on average) reduces retraining times.

\subsection{Edge Inference}
\label{ss:edge}
\vspace{-4pt}

Upon receiving a new set of merged models from \name{}'s cloud component, an edge server quickly incorporates those models into its inference schedule. However, to ensure that merging benefits are maximized, the schedule is altered to reduce the amount of data that must be loaded across the anticipated swaps. During the offline profiling Nexus uses to select per-model batch sizes, \name{} estimates per-workload-iteration swapping delays based on per-model computation costs and swapping delays (both influenced by merging). The idea is that, when merging is used, in addition to ordering models to reduce the number of swaps, models that share the most layers should be placed next to one another in the load order. This lowers the cost of each swap by enabling finer-grained swapping, where only those layers in the next model that are not already in GPU memory must be loaded.

\add{More generally, all schedulers will reap merging benefits in the event that \name{} enables a workload to entirely fit on an edge box (without swapping).
Additional benefits depend on the specific scheduler. For schedulers that employ a statically-configured load order~\cite{nexus,tf_serving}, \name{} can directly modify the schedule as described above to maximize benefits.
Other schedulers~\cite{clockwork} dynamically select the load order to optimize for a certain metric. Such schedulers typically incorporate model loading times when estimating the efficacy of different orders, and thus would naturally factor in the effects of merging per swap. Note that merging benefits would be considered in the context of meeting the optimization metric(s) rather than minimizing global loading delays (as in \name{}'s Nexus variant). Lastly, schedulers that ignore load times in favor of policies such as FIFO~\cite{yarn} or priority scheduling~\cite{slurm} will only see merging-induced reductions in loading costs if merged models are (by chance) neighbors in the order. 
Note that finer-grained~\cite{SwapAdvisor,antman} and space-sharing~\cite{MPS, TensorRT, ekya, mig} schedulers follow the same principles: shared layers should be adjacent in the load orders for the former, while models with the most shared layers should be placed in the same GPU partition for the latter.}
\section{Evaluation}\label{sec:eval}
\vspace{-4pt}
We primarily evaluated \name{} across the diverse workloads and settings from \S\ref{s:workloads}. Our key findings are:

\squishlist
\item \name{} improves per-workload accuracies by \fillin{8-39}\% compared to time/space-sharing strategies alone; these improvements result from \name{} processing \fillin{13-44}\% more frames (while adhering to SLAs).
\item \name{} lowers memory needs by \fillin{17.5-60.7}\% (\fillin{0.2-5.1} GB); savings are \fillin{5.9-52.3}\% more than Mainstream~\cite{MainstreamATC2018} (stem sharing), and within \fillin{9.3-29.0}\% of an optimal that ignores weights (and accuracy drops) when sharing layers.
\item More than 70\% of \name{}'s memory savings are achieved within the first 24-210 minutes of merging+retraining due to its incremental merging heuristic.
\squishend

\subsection{Overall Performance}
\label{ss:overall}

\begin{figure}[!tbp]
  \centering
  \includegraphics[width=0.77\linewidth]{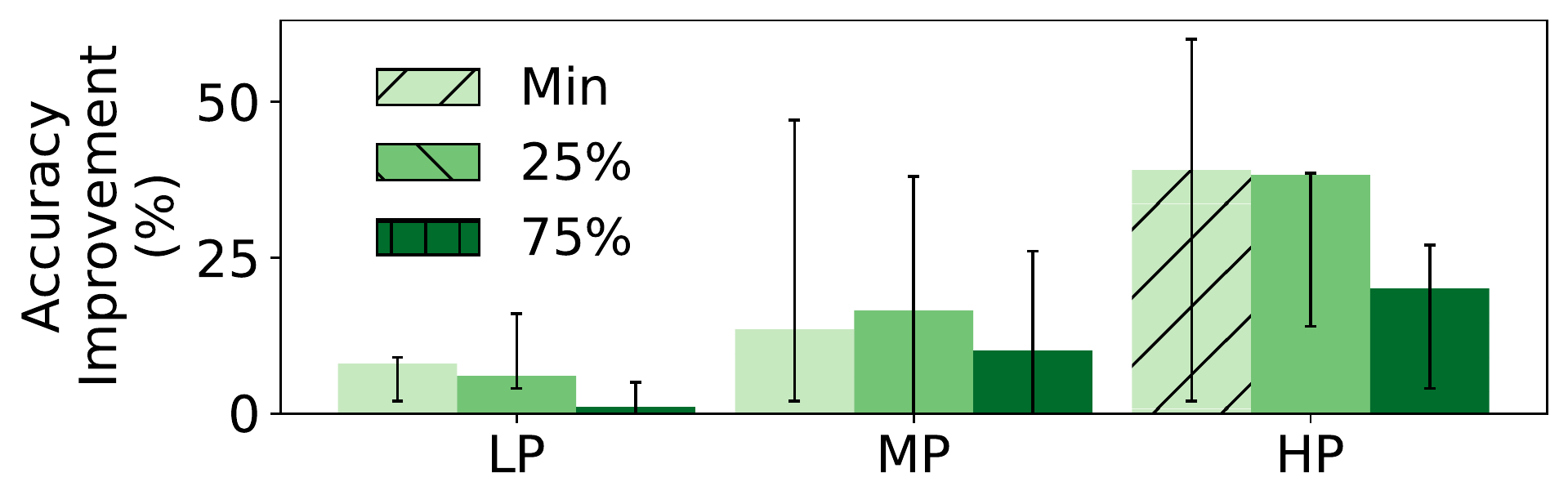}
  \vspace{-10pt}
  \caption{Accuracy improvements with \name{} compared to time/space-sharing alone for different GPU memories (defined in \S\ref{s:workloads}). Bars list median workloads, with error bars as min-max.}
  \label{fig:main_acc}
\end{figure}

\para{End-to-end Accuracy Improvements.} We first compare \name{} with time/space-sharing solutions alone, \ie the Nexus variant running with only unmerged (original) models. Our experiments consider all workloads and resource settings from \S\ref{s:workloads}, a per-frame processing SLA of 100 ms, and an accuracy target of 95\%; trends hold for other accuracy targets and SLAs, which we consider in \S\ref{ss:deepdive}.

Figure~\ref{fig:main_acc} presents our results, showing that \name{} improves accuracy by \fillin{8.0}\%, \fillin{13.5}\%, and \fillin{39.1}\% for the median LP, MP, and HP workloads, respectively, when the edge box GPU's memory is just enough to load and run the largest model in each workload, \ie the \emph{min} setting. The origin of these benefits is \name{}'s ability to reduce the time blocked on swapping delays by \fillin{17.9-84.0}\%, which enables processing on \fillin{13-44}\% more frames than without merging.

Our results highlight two other points. First, \name{}'s benefits are highest for workloads that are most significantly bottlenecked by memory restrictions (and thus loading costs). For instance, workloads \fillin{HP1} and \fillin{LP1} exhibit largely different memory vs. computation profiles: loading costs are \fillin{66}\% of computation costs in the former, but only \fillin{15}\% in the latter. Accordingly, \name{}'s accuracy wins across the available memory settings are \fillin{11-60}\% and \fillin{5-16}\% for workloads \fillin{HP1} and \fillin{LP1}. Second, Figure~\ref{fig:main_acc} shows that, as expected, \name{}'s benefits per workload decrease as the available GPU memory grows, \eg accuracy improvements drop to 17.5\% and 10.2\% for the median MP workload when GPU memory grows to 50\% and 75\% of the total workload memory needs. The reason is straightforward: larger GPU memory yields fewer required swaps without merging.

\begin{figure}[!tbp]
  \centering
  \includegraphics[width=0.86\linewidth]{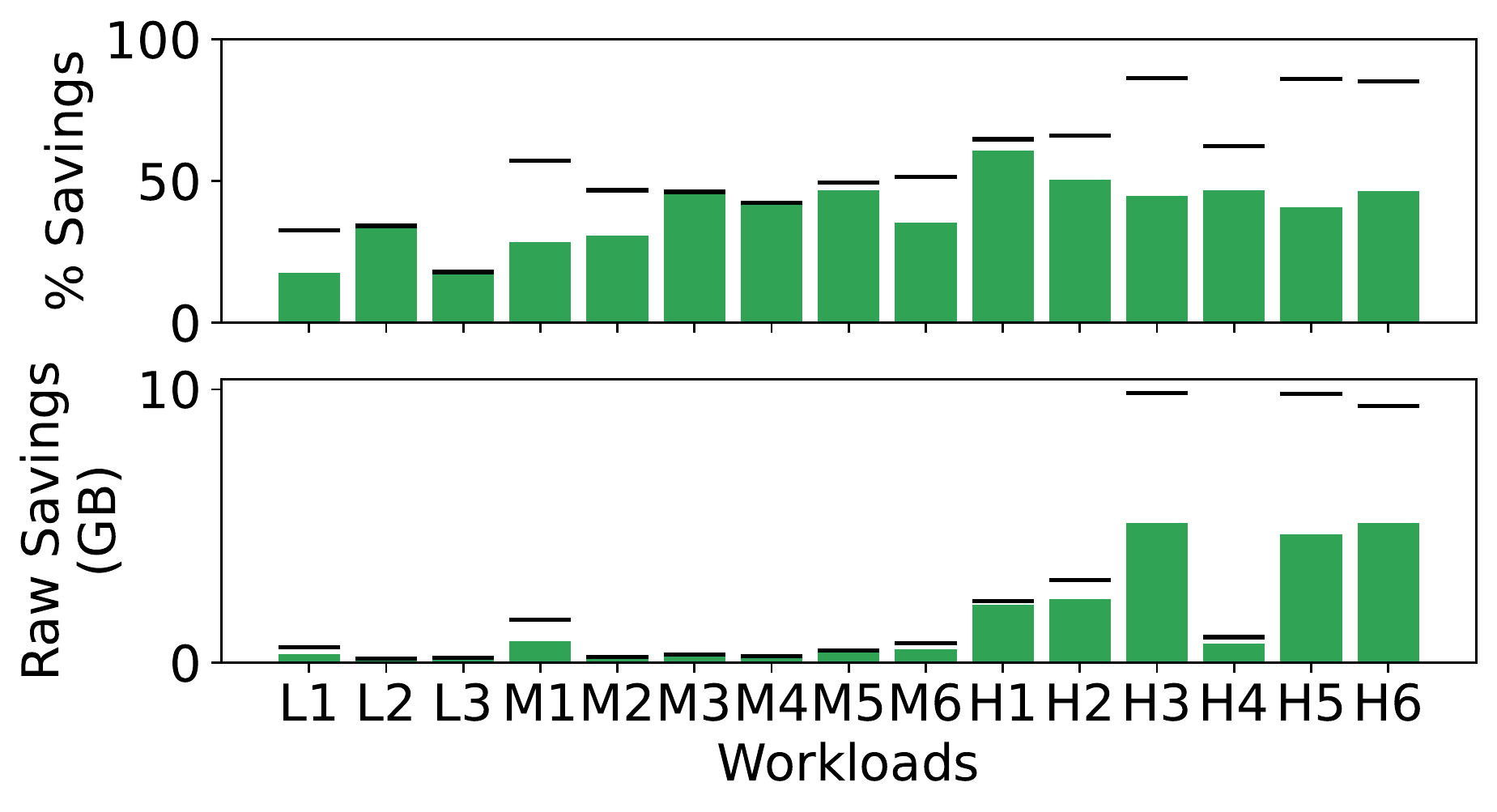}
  \vspace{-10pt}
  \caption{\add{\name{}'s per-workload memory savings. Lines above bars show the theoretical optimal savings from Figure~\ref{fig:upper_memory}.}}
  \vspace{-12pt}
  \label{fig:mem_savings}
\end{figure}

\para{Memory Reductions.} 
Figure~\ref{fig:mem_savings} lists the memory reductions that \name{} delivers for each considered workload by sharing model layers and the associated weights, \ie parameter reductions. We note that reported values here are based on \name{}'s final merging results and an accuracy target of 95\%; we analyze the incremental nature of \name{}'s merging heuristic in \S\ref{ss:deepdive}. As shown, parameter reductions are \fillin{17.5-33.9}\% for LP workloads, \fillin{28.6-46.9}\% for MP workloads, and \fillin{40.9-60.7}\% for HP workloads; the corresponding raw memory savings are \fillin{0.2-0.3} GB, \fillin{0.2-0.8} GB, and \fillin{0.7-5.1} GB, respectively. When analyzed in terms of overall memory usage during inference (\ie including the parameters, inference framework, and intermediate data generated during model execution), reductions are \fillin{4.5-48.1\%} across the workloads. Wins are generally higher for workloads with larger parameter reductions, with the exception of Workloads \fillin{LP1 and LP3} (reductions of 6.3\% and \fillin{4.5}\%) whose intermediates are particularly large relative to the parameters.

To better contextualize the above memory savings, we compare \name{} with two alternatives. First, we consider a theoretical optimal (\emph{Optimal}) that shares all layers that are architecturally identical across a workload's models, without considering accuracy (and the need to find shared weights for those layers). Thus, Optimal represents an \emph{upper bound} on \name{}'s potential memory savings. Second, we compare with \emph{Mainstream}~\cite{MainstreamATC2018}, a recent stem-sharing approach. To run Mainstream, we trained each model in our workloads several times, each time starting with pre-trained weights (based on ImageNet~\cite{ImageNet}) and freezing up to different points, \eg freeze up to layer 10, freeze up to layer 15, etc. We selected the configuration for each model that kept the most layers frozen while meeting the accuracy target (95\% relative to no freezing). Then, within each workload, we merged all layers that were shared across the frozen layer set of the constituent models (note that these layers have identical weights) and
recorded the resultant memory savings. 

\begin{figure}[!tbp]
\centering
  \includegraphics[width=0.77\linewidth]{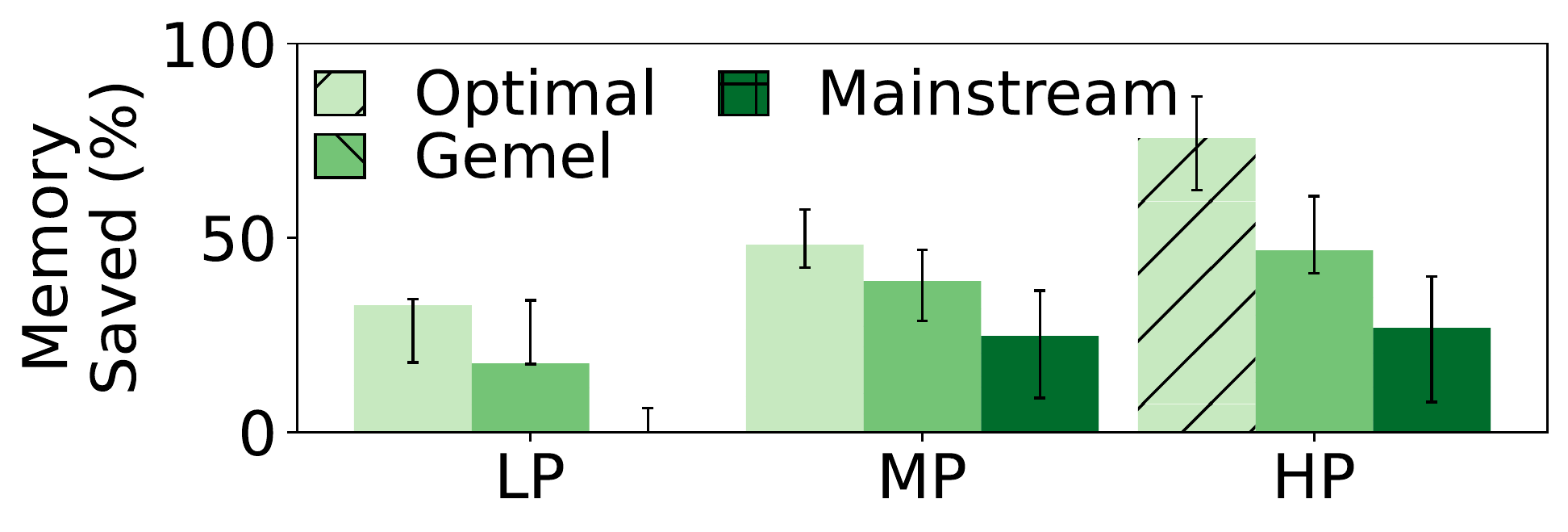}
  \vspace{-10pt}
  \caption{Memory savings with \name{}, an optimal that ignores accuracy, and Mainstream~\cite{MainstreamATC2018}. Bars list the median workload per class, with error bars spanning min to max.}
  \label{fig:mainopt}
\end{figure}

Figure~\ref{fig:mainopt} shows our results, from which we draw two conclusions. 
First, \name{}'s memory savings are within 9.3\%, 15.0\%, and 29.0\% of Optimal for the median LP, MP, and HP workloads. Second, \name{}'s memory reductions are \fillin{5.9-52.3\%} larger than Mainstream's across all workloads. This is a direct consequence of \name{}'s prioritization of memory-heavy layers that routinely appear towards the end of models (\S\ref{ss:observations}). By requiring shared stems from the start of the models, Mainstream would have to share all layers up to the memory-heavy ones; we find that sharing nearly-entire models is rarely possible while meeting accuracy targets (Figure~\ref{fig:sharing_acc}).
The high variance in Mainstream’s results are due to the fact that different models drop in accuracy at different rates when more layers are frozen. 
Classifiers drop relatively slowly (savings up to \fillin{70.1\%}), while detectors are a harder task with faster accuracy drops (Mainstream was unable to share many layers, with savings as low as \fillin{1.0}\%).

\subsection{Analyzing \name{}}
\label{ss:deepdive}
\vspace{-6pt}

\begin{figure}[t]
	\begin{subfigure}[b]{0.5\linewidth}
		\centering
		\includegraphics[width=\textwidth]{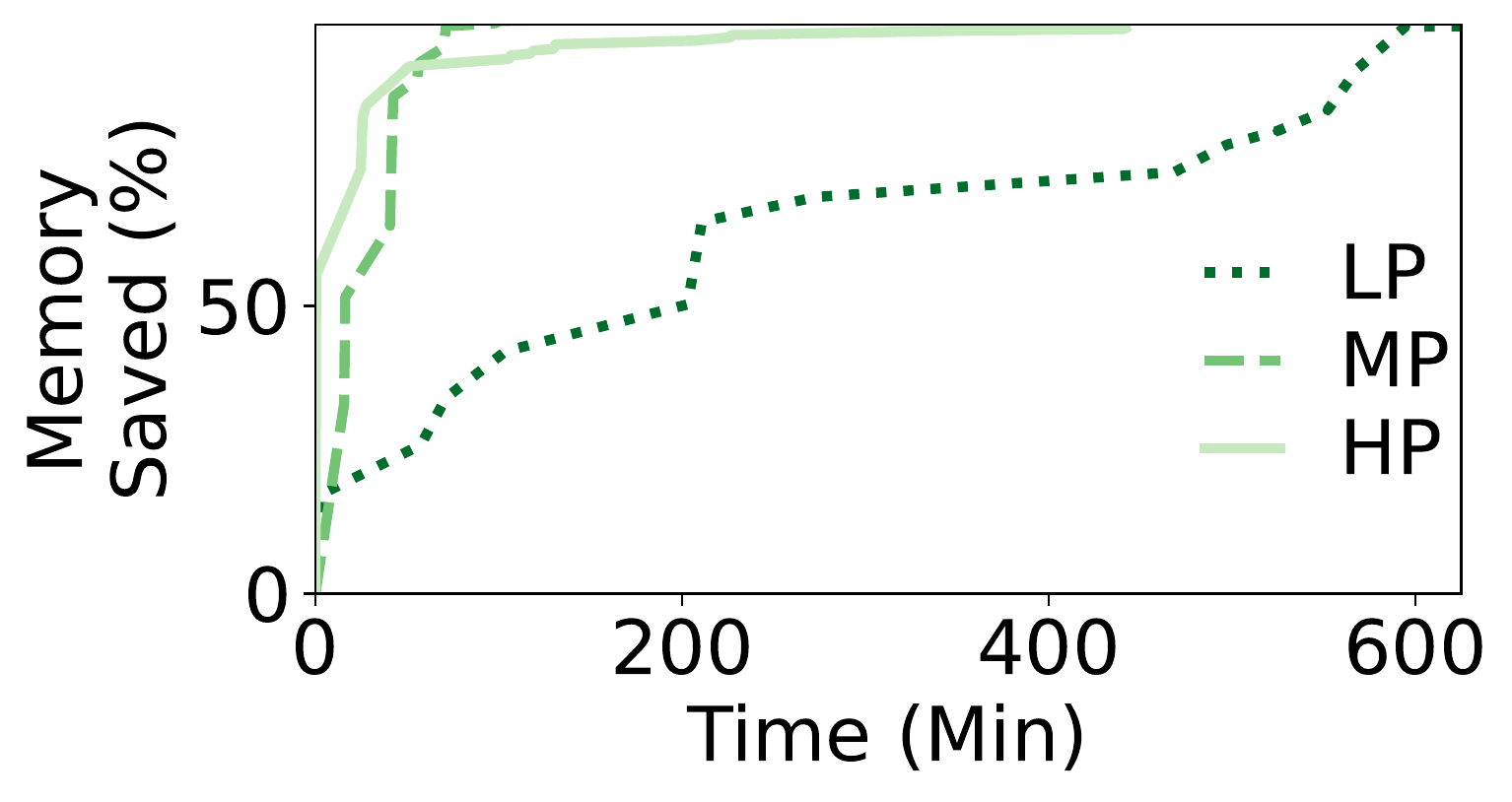}
	\end{subfigure}\hfill
	\begin{subfigure}[b]{0.5\linewidth}
		\centering
		\includegraphics[width=\textwidth]{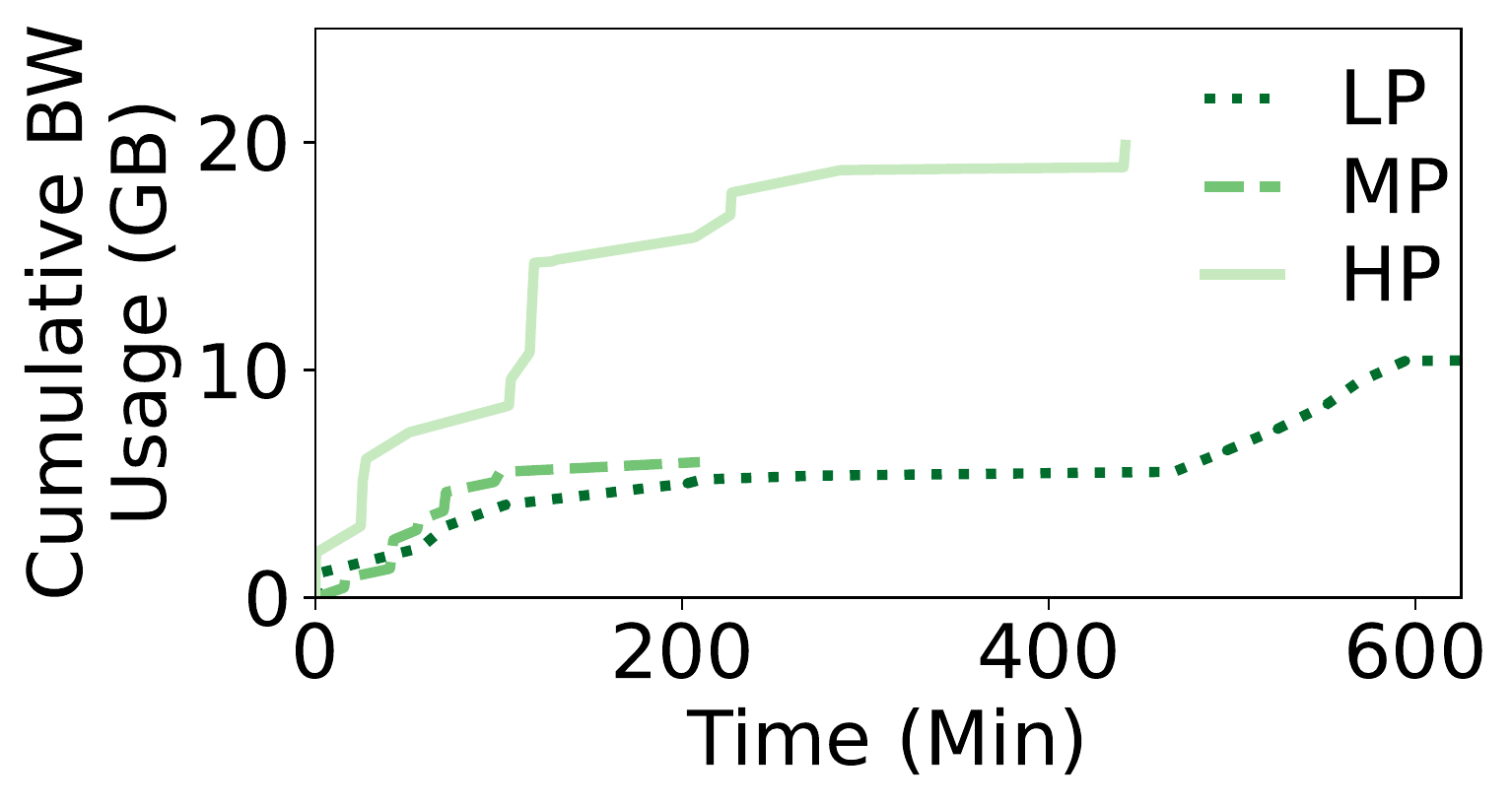}
	\end{subfigure}
	\vspace{-10pt}
	\tightcaption{\name{}'s memory savings (left) and cloud-to-edge bandwidth usage (right) over time during incremental merging. Results show the median workload per class. }
	\vspace{9pt}
	\label{fig:savings_and_bandwidth}
\end{figure}

\para{Incremental memory savings.} Key to \name{}'s practicality are its efficient merging heuristic and retraining optimizations that aim to reap memory savings early in the process; indeed, this is important not only to reap accuracy-friendly memory wins quickly, but also to quickly respond to workload changes.  
As shown in Figure~\ref{fig:savings_and_bandwidth} (left), \fillin{73\%} of \name{}'s achieved memory savings for the median HP workload are realized within the first 24 minutes of merging. Similarly, \fillin{86\% and 64\%} of the total memory savings are achieved in the first 42 and 210 minutes of merging for median MP and LP workloads, respectively.
  
\para{Network bandwidth usage.}  
After each successful merging iteration, \name{} ships weights to edge servers for all updated models.  
As shown in Figure~\ref{fig:savings_and_bandwidth} (right), cumulative bandwidth usage during merging is \fillin{6.0-19.4} GB for the three workloads. Importantly, bandwidth consumption largely grows after substantial memory savings are already reaped. For example, for the median MP workload, 86\% of memory savings are achieved in 42 minutes, while only 2.1 GB (of the total 6.0 GB) of bandwidth is used during that time. The reason is that later merging iterations explore the larger number of lower-memory layers. Thus, \name{} can often deliver large memory savings even in constrained settings with bandwidth caps. 
Note that shipping weights uses cloud-to-edge (not precious edge-to-cloud) bandwidth.

\para{\add{Micro-benchmarks.}} \add{\S\ref{ss:microbenchmarks} profiles the time spent in each of \name{}'s components. Training delays are configurable (Figure~\ref{fig:savings_and_bandwidth}), but dominate cloud merging, with the remaining $<$2\% of time spent on identifying shareable layers and serializing/saving weights from successful training. The majority of time spent at the edge steadily shifts from model loading to inference as \name{}'s incremental merging results stream in; applying results takes $<$.15s and is not blocking.}

\para{Varying accuracy, FPS, and SLA.}
To evaluate the impact of each parameter, we conducted experiments using one randomly selected workload from each class. In each experiment, we only vary one parameter, while keeping the other two at the fixed values from above (95\%, 30 FPS, 100 ms).

Figure~\ref{fig:vary_params} in \ref{ss:additional_figures} presents our results, which exhibit three trends. \add{First, \name{}'s accuracy wins over time/space-sharing alone grow (by \fillin{1.1-7.8}\% for the three workloads) as accuracy targets drop (from 95\% to 80\%)}. This is because certain layers failed to meet 95\% during retraining, but did meet a lower accuracy target. \add{Second, \name{}'s accuracy wins drop as input video frame rates (FPS) drop, \eg from \fillin{6.2-42}\% across the workloads when FPS drops from 30 fps to 5 fps.} The reason is that lower FPS values reduce the amount of inference in any time window (assuming a fixed SLA), which in turn adds tolerance to high loading delays. \add{Third, \name{}'s benefits grow as SLAs become stricter: accuracy wins for the three workloads rise by \fillin{0.4-2.3}\% when SLA drops from 400 to 100 ms.} This is because tighter SLAs imply more skipped frames for a given swapping delay.

\begin{figure}[t]
  \centering
  \includegraphics[width=0.9\linewidth]{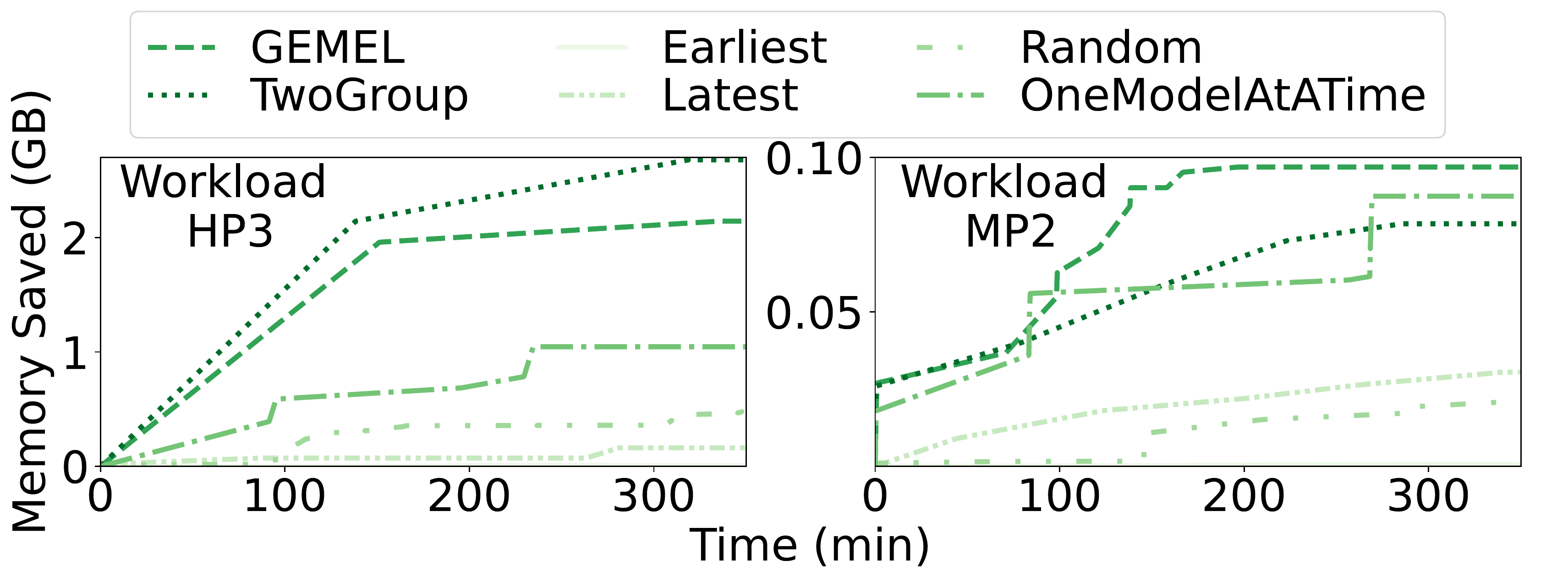}
  \vspace{-10pt}
  \caption{\add{Comparing variants of \name{}'s merging heuristic on two representative workloads.}}
  \vspace{-6pt}
  \label{fig:heuristic_comp}
  \end{figure}
  
\para{\add{Comparison to other merging heuristics.}}
\add{We consider variants that differ from \name{} in one of two ways: they choose layers to merge in a different order or they merge a different number of layers at a time. We describe the variants of each type below, along with the corresponding results. Our experiments use all workloads from \S\ref{s:workloads}, and we report memory saved over time. Figure~\ref{fig:heuristic_comp} shows results for two representative workloads (HP3, MP2); the remaining workload results are in \S\ref{ss:additional_figures}. In summary, no variant consistently outperforms \name{}, and the degradations (in saved memory or merging delays) that each brings to certain workloads (from being overly aggressive or cautious) are substantial.}

\add{Rather than merging layers in descending order of memory usage (irrespective of position) as \name{} does, the variants we consider start by merging the models' earliest layers (\textit{Earliest}), latest layers (\textit{Latest}), and three random layer orderings (\textit{Random}). Across all workloads, these heuristics all resulted in significantly lower memory savings. Among the three, \textit{Latest} performed the best (median of 13.5\% of \name{}'s savings), as memory-heavy layers often appear later in a model (but not necessarily the end). For the same reason, \textit{Earliest} performed the worst (0.2\% of \name{}'s savings). \textit{Random}'s performance varied dramatically (0.2\% - 72.9\%, median of 5.7\% of \name{}'s savings) based on whether a memory-heavy layer was selected.}

\add{We consider two variants to \name{}'s approach of adding one layer group at a time across all models that layer appears in. First, \textit{TwoGroup} more aggressively adds two groups at a time. This can result in faster memory savings than \name{} (3/15 workloads, including Figure~\ref{fig:heuristic_comp} (left)), but most often (8/15 workloads) misses accuracy targets and results in substantial slowdowns (78 min longer to max savings for the median workload). The reason is that, on failure, \textit{TwoGroup} restarts training with 1 group, adding long delay without memory savings, e.g., x=75-220 min in Figure~\ref{fig:heuristic_comp} (right). Second, \emph{OneModelAtATime} less aggressively shares the selected group's layer iteratively across the models it appears in. This reaches within 5\% of \name{}'s memory savings in 8/15 workloads, but is often unnecessarily slow, e.g., in Figure~\ref{fig:heuristic_comp} (left), \name{} successfully considers 5 models at once, while \emph{OneModelAtATime} individually adds models (some of which fail) leading to the flat stretch from 0-91 min.} 

\subsection{\add{Generalization Study}}
\label{ss:generalization}

\add{We evaluate \name{} on over 850 more workloads that extend our main ones by adding: (1) new scene types and the objects they bring (e.g., bags, hats, and people at a beach, boats in a canal), and (2) new models, including more variants in the same families (\eg ResNet, VGG), and entirely new architectures (\eg GoogLeNet~\cite{googlenet}, DenseNet~\cite{densenet}). In total, our analysis involves 17 videos (8 scene types), 13 objects, and 16 models; \S\ref{ss:query_params} lists the values.}

\para{\add{Constructing workloads.}} \add{Each query in a workload is parameterized by a set of knobs: \emph{camera feed} (and corresponding \emph{scene} type), \emph{model}, and \emph{object of interest}. To study the impact of varying each knob (or combination of knobs) on \name{}'s merging, we construct workloads as follows. For each set of target knobs to vary, we start with a random query and incrementally add new queries that only vary values for the target knobs to generate workloads with 2-5 queries each. We did this up to 30 times each for all target knob sets (as their values permit), excluding only (1) target knob sets that vary the \emph{scene} but not \emph{camera} knob, (2) queries for an \emph{object} that never appears in a given \emph{camera} feed, and (3) workloads with no possible memory sharing opportunities.}

\begin{figure}
    \centering
    \includegraphics[width=.86\columnwidth]{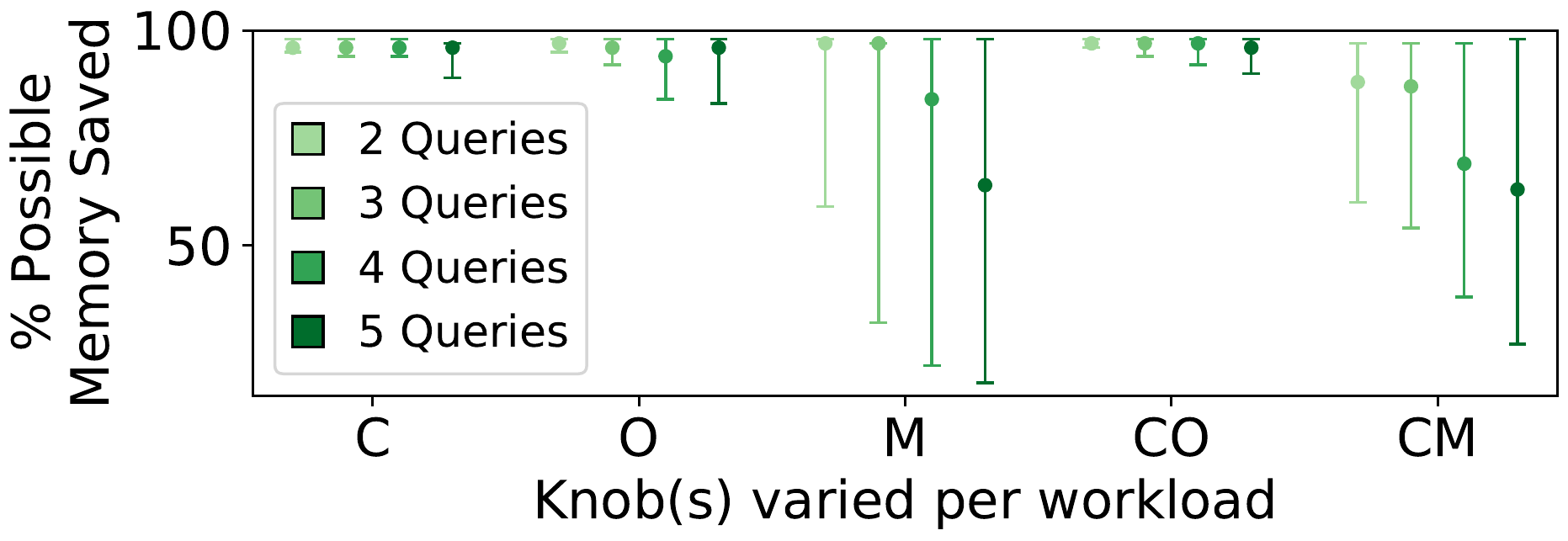}
    \vspace{-7pt}
    \caption{\add{Memory savings across subset of generalization workloads, organized by workload size (color) and knobs varied (\underline{C}amera, \underline{O}bject, \underline{M}odel). Distributions show median and 25-75\%ile; accuracy target was 95\%. 
    Figure~\ref{fig:gen_extensive_appendix} has full graph.}}
    \vspace{-4pt}
    \label{fig:gen_extensive}
\end{figure}

\para{\add{Findings.}} \add{As shown in Figure~\ref{fig:gen_extensive}, \name{}'s memory savings are high for 2-query workloads (89-98\% of optimal at medians), but steadily degrade as workloads grow. This is expected as increasing workload size is (by design, and unrealistically) increasing heterogeneity in this experiment. The nature of degradation depends on the knob(s) being varied. For all combinations of \{\emph{camera, object, scene}\}, degradations are mild moving from 2- to 5-query workloads (0-8\%), showing \name{}'s robustness to variations on those properties. Since \emph{model} is constant in these cases, degradations are because the same set of shareable layers must support more diverse scenarios (making it harder to find shared weights).}

\add{The \emph{Model} knob (alone or with other knobs) presents a different picture, with larger drops in median memory savings (2-33\%) and broader distributions. We can decompose this into two aspects as workload sizes increase:
\squishlist
\item Previously-shared layers appear in the new model: the effect on memory savings heavily depends on where the shared layer appears in the new model; recall that layers can appear in different positions (and thus, serve different roles) across models (Figure~\ref{fig:deep_dive_sharing3}). Cases where the new model introduces drastically different positions for shared layers (e.g., ResNet variants) account for the low-end of the resultant distributions, while memory savings largely persist when positions of shared layer(s) are similar in the new model (e.g., merging across VGG variants). 
\item New layers are shareable with the new model: the extra sharing opportunities increase potential savings, but are more challenging to realize as they reduce the number of non-shared layers whose weights help compensate for the constraints from sharing (\S\ref{ss:challenges}).
\squishend}

\section{\fontsize{11}{6}{\add{Additional Related Work}}}\label{sec:related}
\vspace{-4pt}
\add{
\ref{ss:additional_related} details other related work. Certain systems reuse model components~\cite{sarwar2019incremental, caruana1997multitask, sun2019adashare, vandenhende2019branched}, most relatedly via stem sharing for compute savings~\cite{MainstreamATC2018} or sharing operators with identical weights anywhere in models~\cite{pretzel}; in contrast, \name{} targets memory savings, and enables sharing architecturally-identical layers anywhere in models even if they have different weights. Other platforms optimize model serving either by tuning video analytics-specific knobs to lower compute footprints~\cite{chameleon, VideoStormNSDI2017, FocusOSDI2018, NoScopeVLDB2017, filterforward, dds,cloudseg,vigil,spatula,deep_decision}, or by identifying lightweight variants of individual models that match specific hardware resources~\cite{InFaaSATC21,mcdnn}; \name{} focuses on memory (not compute) bottlenecks, and optimizes across models. Lastly, some frameworks reuse results across frames~\cite{potluck,FreezingHotCloud19,ClipperNSDI17,reducto,glimpse-sensys15}, reducing frame rates for inference and alleviating the impact of model loading delays. \name{} provides benefits at lower fps (\S\ref{ss:deepdive}), and also can alleviate memory pressure across spatially correlated feeds that exhibit limited reuse opportunities at the same time (\S\ref{ss:limitations}).}
\section{\fontsize{11}{9}{Conclusion}}\label{sec:conclusion}
\vspace{-5pt}

Model merging is a new memory management technique that exploits architectural similarities across vision DNNs by sharing their common layers (including parameters but not intermediates). \name{} efficiently carries out model merging by quickly finding and retraining accuracy-preserving layer sharing configurations, and scheduling edge inference to maximize merging benefits (8-39\% accuracy boosts).


\phantomsection
\label{EndOfPaper}
\balance
\Urlmuskip=0mu plus 1mu\relax
\bibliographystyle{abbrv}
\bibliography{references}

\clearpage
\sloppypar
\appendix
\newpage
\twocolumn
\section{Appendix}
\label{s:appendix}
\subsection{Workload Details}
\label{ss:workload_details}

\begin{table}[H]
\footnotesize
\centering
\begin{tabular}{|p{1.3cm}|p{1.4cm}|p{3.9cm}|}
\hline
\textbf{Model} & \textbf{Video Feed}       & \textbf{Object} \\ \hline
frcnn-r101  &A1     & people \\ \hline
r101  &A1     & person, car, bus, truck \\ \hline
r50  &A2     & person, car, bus, truck \\ \hline
r152  &A3     & person, vehicle \\ \hline
mnet  &A4     & person, car, truck \\ \hline
yolo  &A5     & people \\ \hline
tiny-yolo  &A1     & people \\ \hline
ssd-vgg  &A6     & cars \\ \hline
ssd-vgg  &A1     & cars \\ \hline
ssd-mnet  &A5     & cars \\ \hline
ssd-mnet  &A4     & cars \\ \hline
ssd-mnet  &A6     & cars \\ \hline
inception  &A3     & person, vehicle \\ \hline
\end{tabular}
\vspace{-5pt}
\caption{Workload LP1}
\end{table}
\vspace{-15pt}
\begin{table}[H]
\footnotesize
\centering
\begin{tabular}{|p{1.3cm}|p{1.4cm}|p{3.9cm}|}
\hline
\textbf{Model} & \textbf{Video Feed}       & \textbf{Object} \\ \hline
r152  &B1     & person, vehicle \\ \hline
r101  &B2     & person, car, bus, truck \\ \hline
ssd-vgg  &B3     & people \\ \hline
\end{tabular}
\vspace{-5pt}
\caption{Workload LP2}
\end{table}
\vspace{-15pt}
\begin{table}[H]
\footnotesize
\centering
\begin{tabular}{|p{1.3cm}|p{1.4cm}|p{3.9cm}|}
\hline
\textbf{Model} & \textbf{Video Feed}       & \textbf{Object} \\ \hline
ssd-mnet  &B4     & cars \\ \hline
frcnn-r101  &B3     & people \\ \hline
r152  &B1     & person, vehicle \\ \hline
r18  &B3     & person, car, bus, truck, motorbike \\ \hline
inception  &B1     & person, vehicle \\ \hline
\end{tabular}
\vspace{-5pt}
\caption{Workload LP3}
\end{table}
\vspace{-15pt}
\begin{table}[H]
\footnotesize
\centering
\begin{tabular}{|p{1.3cm}|p{1.4cm}|p{3.9cm}|}
\hline
\textbf{Model} & \textbf{Video Feed}       & \textbf{Object} \\ \hline
frcnn-r50  &B1     & cars \\ \hline
frcnn-r50  &B1     & people \\ \hline
r50  &B2     & person, car, bus, truck \\ \hline
r50  &B1     & person, vehicle \\ \hline
r152  &B3     & person, car, bus, truck, motorbike \\ \hline
r152  &B4     & person, car, bus, truck \\ \hline
r18  &B5     & person, car, bus, truck \\ \hline
r18  &B4     & person, car, bus, truck \\ \hline
tiny-yolo  &B3     & cars \\ \hline
tiny-yolo  &B2     & cars \\ \hline
yolo  &B5     & cars \\ \hline
yolo  &B1     & cars \\ \hline
ssd-vgg  &B4     & cars \\ \hline
ssd-vgg  &B3     & people \\ \hline
inception  &B3     & person, car, bus, truck, motorbike \\ \hline
\end{tabular}
\vspace{-5pt}
\caption{Workload MP1}
\end{table}
\vspace{-15pt}

\begin{table}[H]
\footnotesize
\centering
\begin{tabular}{|p{1.3cm}|p{1.4cm}|p{3.9cm}|}
\hline
\textbf{Model} & \textbf{Video Feed}       & \textbf{Object} \\ \hline
r50  &B3     & person, car, bus, truck, motorbike \\ \hline
r50  &B1     & person, vehicle \\ \hline
r152  &B3     & person, car, bus, truck, motorbike \\ \hline
r18  &B5     & person, car, bus, truck \\ \hline
ssd-mnet  &B1     & cars \\ \hline
ssd-mnet  &B2     & cars \\ \hline
\end{tabular}
\vspace{-5pt}
\caption{Workload MP2}
\end{table}
\vspace{-15pt}
\begin{table}[H]
\footnotesize

\centering
\begin{tabular}{|p{1.3cm}|p{1.4cm}|p{3.9cm}|}
\hline
\textbf{Model} & \textbf{Video Feed}       & \textbf{Object} \\ \hline
yolo  &B4     & cars \\ \hline
yolo  &B3     & people \\ \hline
tiny-yolo  &B1     & people \\ \hline
tiny-yolo  &B3     & cars \\ \hline
ssd-vgg  &B1     & cars \\ \hline
ssd-mnet  &B5     & cars \\ \hline
\end{tabular}
\vspace{-5pt}
\caption{Workload MP3}
\end{table}
\vspace{-15pt}
\begin{table}[H]
\footnotesize
\centering
\begin{tabular}{|p{1.3cm}|p{1.4cm}|p{3.9cm}|}
\hline
\textbf{Model} & \textbf{Video Feed}       & \textbf{Object} \\ \hline
yolo  &A4     & people \\ \hline
yolo  &A6     & cars \\ \hline
r50  &A2     & person, car, bus, truck \\ \hline
\end{tabular}
\vspace{-5pt}
\caption{Workload MP4}
\end{table}
\vspace{-15pt}
\begin{table}[H]
\footnotesize
\centering
\begin{tabular}{|p{1.3cm}|p{1.4cm}|p{3.9cm}|}
\hline
\textbf{Model} & \textbf{Video Feed}       & \textbf{Object} \\ \hline
yolo  &A5     & people \\ \hline
yolo  &A4     & people \\ \hline
r152  &A1     & person, car, bus, truck \\ \hline
r152  &A4     & person, car, truck \\ \hline
mnet  &A4     & person, car, truck \\ \hline
\end{tabular}
\vspace{-5pt}
\caption{Workload MP5}
\end{table}
\vspace{-15pt}
\begin{table}[H]
\footnotesize
\centering
\begin{tabular}{|p{1.3cm}|p{1.4cm}|p{3.9cm}|}
\hline
\textbf{Model} & \textbf{Video Feed}       & \textbf{Object} \\ \hline
frcnn-r50  &B5     & cars \\ \hline
frcnn-r50  &B4     & cars \\ \hline
r50  &B2     & person, car, bus, truck \\ \hline
mnet  &B3     & person, car, bus, truck, motorbike \\ \hline
tiny-yolo  &B3     & people \\ \hline
\end{tabular}
\vspace{-5pt}
\caption{Workload MP6}
\end{table}
\vspace{-15pt}
\begin{table}[H]
\footnotesize
\centering
\begin{tabular}{|p{1.3cm}|p{1.4cm}|p{3.9cm}|}
\hline
\textbf{Model} & \textbf{Video Feed}       & \textbf{Object} \\ \hline
vgg  &B4     & person, car, bus, truck \\ \hline
vgg  &B1     & person, vehicle \\ \hline
vgg  &B3     & person, car, bus, truck, motorbike \\ \hline
vgg  &B5     & person, car, bus, truck \\ \hline
ssd-vgg  &B5     & cars \\ \hline
ssd-mnet  &B5     & cars \\ \hline
mnet  &B4     & person, car, bus, truck \\ \hline
tiny-yolo  &B3     & cars \\ \hline
tiny-yolo  &B1     & people \\ \hline
frcnn-r50  &B4     & cars \\ \hline
frcnn-r50  &B5     & cars \\ \hline
\end{tabular}
\vspace{-5pt}
\caption{Workload HP1}
\end{table}

\vspace{-15pt}
\begin{table}[H]
\footnotesize
\centering
\begin{tabular}{|p{1.3cm}|p{1.4cm}|p{3.9cm}|}
\hline
\textbf{Model} & \textbf{Video Feed}       & \textbf{Object} \\ \hline
frcnn-r101  &B4     & cars \\ \hline
frcnn-r101  &B5     & cars \\ \hline
frcnn-r101  &B1     & cars \\ \hline
frcnn-r101  &B2     & cars \\ \hline
frcnn-r50  &B1     & people \\ \hline
r50  &B3     & person, car, bus, truck, motorbike \\ \hline
r18  &B3     & person, car, bus, truck, motorbike \\ \hline
ssd-mnet  &B3     & people \\ \hline
ssd-mnet  &B1     & people \\ \hline
mnet  &B4     & person, car, bus, truck \\ \hline
yolo  &B3     & people \\ \hline
tiny-yolo  &B5     & cars \\ \hline
tiny-yolo  &B1     & people \\ \hline
vgg  &B4     & person, car, bus, truck \\ \hline
inception  &B2     & person, car, bus, truck \\ \hline
inception  &B3     & person, car, bus, truck, motorbike \\ \hline
\end{tabular}
\vspace{-5pt}
\caption{Workload HP2}
\end{table}
\vspace{-15pt}
\begin{table}[H]
\footnotesize
\centering
\begin{tabular}{|p{1.3cm}|p{1.4cm}|p{3.9cm}|}
\hline
\textbf{Model} & \textbf{Video Feed}       & \textbf{Object} \\ \hline
frcnn-r50  &A3     & cars \\ \hline
frcnn-r50  &A3     & people \\ \hline
frcnn-r50  &A1     & cars \\ \hline
frcnn-r50  &A1     & people \\ \hline
frcnn-r50  &A5     & cars \\ \hline
frcnn-r50  &A5     & people \\ \hline
frcnn-r50  &A2     & cars \\ \hline
frcnn-r50  &A4     & cars \\ \hline
frcnn-r50  &A2     & trucks \\ \hline
frcnn-r101  &A3     & people \\ \hline
yolo  &A3     & cars \\ \hline
yolo  &A3     & people \\ \hline
yolo  &A1     & people \\ \hline
yolo  &A7     & buses \\ \hline
yolo  &A7     & cars \\ \hline
yolo  &A7     & people \\ \hline
yolo  &A7     & trucks \\ \hline
yolo  &A5     & trucks \\ \hline
yolo  &A5     & people \\ \hline
yolo  &A6     & cars \\ \hline
r152  &A3     & person, vehicle \\ \hline
r152  &A1     & person, car, bus, truck \\ \hline
r152  &A7     & person, car, bus, truck \\ \hline
r152  &A6     & car, bus, truck \\ \hline
r152  &A2     & person, car, bus, truck \\ \hline
r152  &A4     & person, car, truck \\ \hline
r50  &A3     & person, vehicle \\ \hline
r50  &A7     & person, car, bus, truck \\ \hline
r50  &A6     & car, bus, truck \\ \hline
r50  &A2     & person, car, bus, truck \\ \hline
r50  &A6     & car, bus, truck \\ \hline
ssd-vgg  &A3     & people \\ \hline
ssd-vgg  &A1     & cars \\ \hline
ssd-vgg  &A5     & people \\ \hline
ssd-vgg  &A6     & cars \\ \hline
ssd-vgg  &A4     & cars \\ \hline
vgg  &A2     & person, car, bus, truck \\ \hline
r18  &A2     & person, car, bus, truck \\ \hline
\end{tabular}
\vspace{-5pt}
\caption{Workload HP3}
\end{table}
\vspace{-15pt}
\begin{table}[H]
\footnotesize
\centering
\begin{tabular}{|p{1.3cm}|p{1.4cm}|p{3.9cm}|}
\hline
\textbf{Model} & \textbf{Video Feed}       & \textbf{Object} \\ \hline
yolo  &B1     & cars \\ \hline
yolo  &B5     & cars \\ \hline
tiny-yolo  &B2     & cars \\ \hline
tiny-yolo  &B1     & cars \\ \hline
tiny-yolo  &B3     & people \\ \hline
ssd-vgg  &B5     & cars \\ \hline
ssd-vgg  &B3     & people \\ \hline
ssd-mnet  &B5     & cars \\ \hline
ssd-mnet  &B3     & people \\ \hline
ssd-mnet  &B2     & cars \\ \hline
ssd-mnet  &B1     & people \\ \hline
mnet  &B3     & person, car, bus, truck, motorbike \\ \hline
mnet  &B5     & person, car, bus, truck \\ \hline
r152  &B4     & person, car, bus, truck \\ \hline
r152  &B3     & person, car, bus, truck, motorbike \\ \hline
r152  &B1     & person, vehicle \\ \hline
\end{tabular}
\vspace{-5pt}
\caption{Workload HP4}
\end{table}
\vspace{-15pt}
\begin{table}[H]
\footnotesize
\centering
\begin{tabular}{|p{1.3cm}|p{1.4cm}|p{3.9cm}|}
\hline
\textbf{Model} & \textbf{Video Feed}       & \textbf{Object} \\ \hline
frcnn-r50  &A3     & cars \\ \hline
frcnn-r50  &A3     & people \\ \hline
frcnn-r50  &A1     & cars \\ \hline
frcnn-r50  &A1     & people \\ \hline
frcnn-r50  &A5     & cars \\ \hline
frcnn-r50  &A5     & people \\ \hline
frcnn-r50  &A2     & cars \\ \hline
frcnn-r50  &A4     & cars \\ \hline
frcnn-r50  &A2     & trucks \\ \hline
frcnn-r101  &A3     & people \\ \hline
yolo  &A3     & cars \\ \hline
yolo  &A3     & people \\ \hline
yolo  &A1     & people \\ \hline
yolo  &A7     & buses \\ \hline
yolo  &A7     & cars \\ \hline
yolo  &A7     & people \\ \hline
yolo  &A7     & trucks \\ \hline
yolo  &A5     & trucks \\ \hline
yolo  &A5     & people \\ \hline
yolo  &A6     & cars \\ \hline
r152  &A3     & person, vehicle \\ \hline
r152  &A1     & person, car, bus, truck \\ \hline
r152  &A7     & person, car, bus, truck \\ \hline
r152  &A6     & car, bus, truck \\ \hline
r152  &A2     & person, car, bus, truck \\ \hline
r152  &A4     & person, car, truck \\ \hline
r50  &A3     & person, vehicle \\ \hline
r50  &A7     & person, car, bus, truck \\ \hline
r50  &A6     & car, bus, truck \\ \hline
r50  &A2     & person, car, bus, truck \\ \hline
r50  &A6     & car, bus, truck \\ \hline
ssd-vgg  &A3     & people \\ \hline
inception  &A3     & person, vehicle \\ \hline
inception  &A1     & person, car, bus, truck \\ \hline
inception  &A7     & person, car, bus, truck \\ \hline
inception  &A6     & car, bus, truck \\ \hline
inception  &A4     & person, car, truck \\ \hline
vgg  &A2     & person, car, bus, truck \\ \hline
r18  &A2     & person, car, bus, truck \\ \hline
r18  &A2     & person, car, bus, truck \\ \hline
r18  &A2     & person, car, bus, truck \\ \hline
\end{tabular}
\vspace{-5pt}
\caption{Workload HP5}
\end{table}
\vspace{-15pt}

\begin{table}[H]
\footnotesize
\centering
\begin{tabular}{|p{1.3cm}|p{1.4cm}|p{3.9cm}|}
\hline
\textbf{Model} & \textbf{Video Feed}       & \textbf{Object} \\ \hline
frcnn-r50  &A3     & cars \\ \hline
frcnn-r50  &A3     & people \\ \hline
frcnn-r50  &A1     & cars \\ \hline
frcnn-r50  &A1     & people \\ \hline
frcnn-r50  &A5     & cars \\ \hline
frcnn-r50  &A5     & people \\ \hline
frcnn-r50  &A2     & cars \\ \hline
frcnn-r50  &A4     & cars \\ \hline
frcnn-r50  &A2     & trucks \\ \hline
frcnn-r101  &A3     & people \\ \hline
yolo  &A3     & cars \\ \hline
yolo  &A3     & people \\ \hline
yolo  &A1     & people \\ \hline
yolo  &A7     & buses \\ \hline
yolo  &A7     & cars \\ \hline
yolo  &A7     & people \\ \hline
r101  &A1     & person, car, bus, truck \\ \hline
r101  &A7     & person, car, bus, truck \\ \hline
r101  &A6     & car, bus, truck \\ \hline
\end{tabular}
\vspace{-5pt}
\caption{Workload HP6} 
\end{table}

\begin{table}[H]
\footnotesize
\centering
\begin{tabular}{|p{1.3cm}|p{1.4cm}|p{3.9cm}|}
\hline
\textbf{Model} & \textbf{Video Feed}       & \textbf{Object} \\ \hline
r101  &A1     & person, car, bus, truck \\ \hline
r152  &A3     & person, vehicle \\ \hline
r152  &A1     & person, car, bus, truck \\ \hline
r152  &A7     & person, car, bus, truck \\ \hline
r152  &A6     & car, bus, truck \\ \hline
r152  &A2     & person, car, bus, truck \\ \hline
r152  &A4     & person, car, truck \\ \hline
r50  &A3     & person, vehicle \\ \hline
r50  &A7     & person, car, bus, truck \\ \hline
r50  &A6     & car, bus, truck \\ \hline
r50  &A2     & person, car, bus, truck \\ \hline
r50  &A6     & car, bus, truck \\ \hline
tiny-yolo  &A1     & people \\ \hline
tiny-yolo  &A5     & people \\ \hline
inception  &A3     & person, vehicle \\ \hline
inception  &A1     & person, car, bus, truck \\ \hline
inception  &A7     & person, car, bus, truck \\ \hline
inception  &A6     & car, bus, truck \\ \hline
inception  &A4     & person, car, truck \\ \hline
vgg  &A2     & person, car, bus, truck \\ \hline
r18  &A2     & person, car, bus, truck \\ \hline
r18  &A2     & person, car, bus, truck \\ \hline
r18  &A2     & person, car, bus, truck \\ \hline
\end{tabular}
\vspace{-5pt}
\caption{Workload HP6 (continued)} 
\end{table}

\vspace{-5pt}
\subsection{\add{Generalization Workload Query Knobs}}
\label{ss:query_params}
\vspace{-10pt}
\begin{table}[H]
\centering
\footnotesize
\begin{tabular}{|p{.8cm} | p{6.8cm}|}
\hline
\textbf{Knob} & \textbf{Values} \\ 
\hline
Object & Truck, Person, Bus, Boat, Shoe, Skateboard, Car, Hat, Backpack, Wine Glass, Traffic Light, Parking Meter, Surfboard \\
\hline
Camera & A0, A1, A2, A3, B0, B1, B2, B3, B4, B5, B6, Restaurant, Mall, Beach, Canal, Parking Lot, Street \\
\hline
Model & SSD-VGG, AlexNet, YOLOv3, Tiny-YOLOv3, DenseNet, SqueezeNet, GoogLeNet, ResNet-18, ResNet-34, ResNet-50, ResNet-101, ResNet-152, VGG-11, VGG-13, VGG-16, VGG-19 \\
\hline
Scene & CityA Traffic, CityB Traffic, Restaurant, Beach, Mall, Canal, Parking Lot, Street\\
\hline
\end{tabular}
\vspace{-3pt}
\caption{\add{Knob values considered in generalization study.}}
\label{appendix:gen_query_params}
\end{table}

\vspace{-5pt}
\subsection{\add{Workload Memory Settings}}
\label{ss:memory_settings}
\vspace{-10pt}
\begin{table}[H]
\footnotesize
\centering
\begin{tabular}{|l|l|l|l|} 
\hline
\textbf{Workload} & \textbf{L1} & \textbf{L2} & \textbf{L3} \\\hline
\textbf{Min} & 4.50 & 1.45 & 4.50  \\ \hline 
\textbf{50\%} & 5.12 & 1.59 & 4.72  \\ \hline
\textbf{75\%} & 5.43 & 1.66 & 4.83  \\ \hline
\end{tabular}
\vspace{-5pt}
\caption{\add{Edge box memory settings for LP workloads (in GB).}}
\end{table}
\vspace{-15pt}
\begin{table}[H]
\footnotesize
\centering
\begin{tabular}{|l|l|l|l|l|l|l|} 
\hline
\textbf{Workload} & \textbf{M1} & \textbf{M2} & \textbf{M3} & \textbf{M4} & \textbf{M5} & \textbf{M6} \\\hline
\textbf{Min} & 3.35 & 1.45 & 1.32 & 1.32 & 1.45 & 3.35 \\ \hline 
\textbf{50\%} & 4.56 & 1.62 & 1.55 & 1.45 & 1.83 & 3.77 \\ \hline 
\textbf{75\%} & 5.16 & 1.70 & 1.65 & 1.52 & 2.02 & 3.99 \\ \hline 
\end{tabular}
\vspace{-5pt}
\caption{\add{Edge box memory settings for MP workloads (in GB).}}
\end{table}
\vspace{-15pt}
\begin{table}[H]
\footnotesize
\centering
\begin{tabular}{|l|l|l|l|l|l|l|} 
\hline
\textbf{Workload} & \textbf{H1} & \textbf{H2} & \textbf{H3} & \textbf{H4} & \textbf{H5} & \textbf{H6} \\\hline
\textbf{Min} & 3.35 & 4.50 & 4.50 & 1.45 & 4.50 & 4.50 \\ \hline 
\textbf{50\%} & 4.87 & 6.60 & 10.25 & 2.17 & 10.41 & 10.26 \\ \hline 
\textbf{75\%} & 5.63 & 7.66 & 13.13 & 2.53 & 13.36 & 13.14 \\ \hline 
\end{tabular}
\vspace{-5pt}
\caption{\add{Edge box memory settings for HP workloads (in GB).}}
\end{table}

\vspace{-5pt}
\subsection{\add{Implementation Details}}
\label{ss:implementation_details}
\add{\name{}'s main components are training models at the cloud server and running the scheduler at the edge. During training, a single optimizer manages the weights across all considered models; the optimizer holds a single copy of weights for each layer that is shared across the models. Aside from this, \name{}'s training process mirrors classic multi-task training\cite{caruana1997multitask}: it forms a collective pool of an equal number of data samples from all models and randomly selects batches from this pool. 
Samples are run through their respective models, each model calculates its loss individually, and losses are summed over all models. In this way, layers that are shared are updated by the concurrent training of multiple models within a single batch.}

\add{The Nexus-variant scheduler chooses when to load and evict models as described in \S\ref{ss:edge}. To load a model into GPU memory, the scheduler  simply calls ``.cuda()" on that model's PyTorch object. PyTorch automatically only loads layer weights not already in GPU memory. However, when evicting a model, PyTorch, by default, removes all of the layers' weights from GPU memory. This poses a problem if some of those weights are needed by models still in GPU memory (\ie they are shared). To avoid this, the scheduler: (1) maintains a running list of shared layers that are needed by models currently in GPU memory or next in line to be loaded, and (2) when a model needs to be evicted, only evicts weights corresponding to layers not in the list. Overall, \name{} is implemented in $\approx$3500 LOC: 500 for finding shared layers and sharing them according to the heuristic, 2500 for dataset management and retraining, and 500 for scheduling models at the edge.}

\subsection{\add{Micro-benchmarks}}
\label{ss:microbenchmarks}

\add{We break down time spent in each component of \name{}'s end-to-end system for our main evaluation workloads.
\squishlist
\item \textit{Merging.} The total time spent on merging is configurable, with Figure~\ref{fig:savings_and_bandwidth} illustrating the memory savings that \name{} reaps over time for different workloads. Regardless, training dominates the time spent in the overall merging process. Other tasks include (1) finding shared layers, which is done once per workload and takes 0.7-1.4s, or $<$0.01\% of merging time (we consider a merging time of 6 hours as across workloads, most merging happens by this time), and (2) serializing and saving (to disk) weights after each successful round of training, which takes 9.1-19.5s each time (total of 0.8\%-1.44\% 
of merging time).
\item \textit{Edge inference.} When using the Nexus variant without any merged models, the time spent 
blocked while waiting for models to load is 32.8\%, 48.3\%, and 52.0\% at the median 
of LP, MP, and HP workloads. This time steadily reduces as incremental merging results arrive from \name{}'s cloud merging component, and culminates at 22.1\%, 34.6\%, and 27.9\%
when the final merging results arrive. 
When new merging results arrive, depending on the model size, it takes between 0.03 and 0.58s to load each set of weights into their respective model. 
However, this time is not blocking, as \name{} creates a new version of each affected model in CPU with the up-to-date sharing and weights, and substitutes it into the schedule at the next time the corresponding old model is evicted. Therefore, swapping the new model for the old one does not incur any delay.
\squishend}

\subsection{\add{Additional Related Work}}
\label{ss:additional_related}

\para{Model Sharing.} Sharing parts of models has been explored in the ML literature~\cite{sarwar2019incremental}, and more recently in video analytics through Mainstream~\cite{MainstreamATC2018}, which aims to share outputs from common \emph{stems} of early layers across models. However, unlike \name{} which addresses memory bottlenecks, the main goal of these works is to reduce computation. This leads to two limitations for our problem. First, these approaches only apply to models that operate on the same underlying data, limiting their applicability to realistic workloads with many videos. Second, and more importantly, because memory-heavy layers are often towards the end of vision DNNs (\S\ref{ss:observations}), stem-sharing approaches would have to share almost all model layers to reap large memory savings -- doing so almost always comes with accuracy violations (\S\ref{ss:challenges} and \S\ref{ss:overall}). In contrast, by only sharing weights (not intermediates), \name{} is able to quickly share only late-stage layers that enable memory savings and accuracy preservation.

PRETZEL~\cite{pretzel} focuses on reusing operators in \emph{non-deep} ML models, such as featurizers. The key observation PRETZEL makes is that \emph{both the operator and the parameters are shared} across models, and hence storing them in a shared object store for reuse leads to better memory utilization; the vast majority of savings 
arise from sharing parameters. 
However, in edge video analytics, the assumption that shared layers across models have the same parameters (weights) is often violated, as it is common for vision DNNs
to be specialized to diverse tasks, objects, and videos (\S\ref{ss:challenges}). Indeed, the core challenges that \name{} tackles are in (quickly) determining which architecturally identical layers consume substantial memory \emph{and} can be retrained to use unified weights without violating accuracy requirements.

Multi-task learning is a well-known technique in machine learning that can learn multiple \emph{related} tasks simultaneously~\cite{caruana1997multitask}. Some works also study how to share layers in multi-task learning\cite{sun2019adashare, vandenhende2019branched}. However, the problem is usually studied in the context of transfer learning, where one model does not have enough data to train on, and instead is trained together with another model; this is in contrast to our setting which considers two sets of pretrained weights that must be shared. Additionally, the related tasks are usually variations of the same model (\eg spam classification) and thus the data for each individual task can be pooled together. In contrast, objectives vary across edge video analytics DNNs, \eg detection vs. classification, different objects.

\para{Optimized Video Analytics Pipelines.} 
\add{Many systems aim to lower the resource usage (\eg computation and bandwidth) of video analytics pipelines. }Chameleon~\cite{chameleon} profiles pipeline knobs to identify computationally cheaper configurations that preserve accuracy. VideoStorm~\cite{VideoStormNSDI2017} proposes scheduling techniques that leverage lag tolerance to optimize analytics results. NoScope~\cite{NoScopeVLDB2017} and Focus~\cite{FocusOSDI2018} build support for low-latency queries on large scale video streams. These systems are complementary to \name{}, which focuses on alleviating memory (not compute) bottlenecks in edge video analytics. \add{Other systems partially process frames at the edge to reduce both computation and bandwidth costs in video analytics pipelines~\cite{dds,cloudseg,vigil,filterforward,spatula, deep_decision}. Unlike these systems, \name{} runs inference entirely at the edge.}

\add{Previous work has also explored optimizing model serving systems to reduce computation over large and heterogeneous workloads. MCDNN~\cite{mcdnn} and INFaaS~\cite{InFaaSATC21} generate model variants with a range of resource profiles and when running inference, dynamically choose which variant to run. While MCDNN generates these variants by compressing models (i.e., retraining), InFaaS uses methods such as pruning and quantization. These methods optimize each model individually; \name{} supports such variants and provides additional benefits by optimizing \emph{across} models.}

\para{\add{Result Reuse.}} \add{Other systems reduce required model inference by reusing previously computed results, within a query~\cite{reducto, glimpse-sensys15, FreezingHotCloud19} and across queries~\cite{ClipperNSDI17, potluck}. Within a query, these systems filter frames that are similar enough to a previously computed frame and reuse results from the previous computation~\cite{reducto,glimpse-sensys15}. Across queries, they reuse results when the models and inputs are both the same~\cite{ClipperNSDI17} or the models are the same and the inputs are similar~\cite{potluck}. Less inference leads to a higher tolerance for loading delays, so these systems can alleviate memory pressure, as described in ~\S\ref{ss:deepdive}. However, these methods highly depend on how much inference can be avoided, and with spatially correlated inputs (like \name{}'s), all queries typically require high inference loads at the same time (e.g., dynamic, busier scenes). Therefore, reusing results is insufficient to address the memory bottleneck. These approaches could be combined with \name{}, allowing for lower loading costs and when possible, higher tolerance for those loading costs.} 

\add{Finally, there exist training optimizations that trade off memory usage for computation overheads~\cite{monet,zeroinf,capuchin}. We eschew such techniques given the holistic constraint on compute resources that edge boxes face (\S\ref{s:intro}).}

\subsection{\add{Additional Figures}}
\label{ss:additional_figures}
\begin{figure*}[h!]
  \centering
  \includegraphics[width=0.9\linewidth]{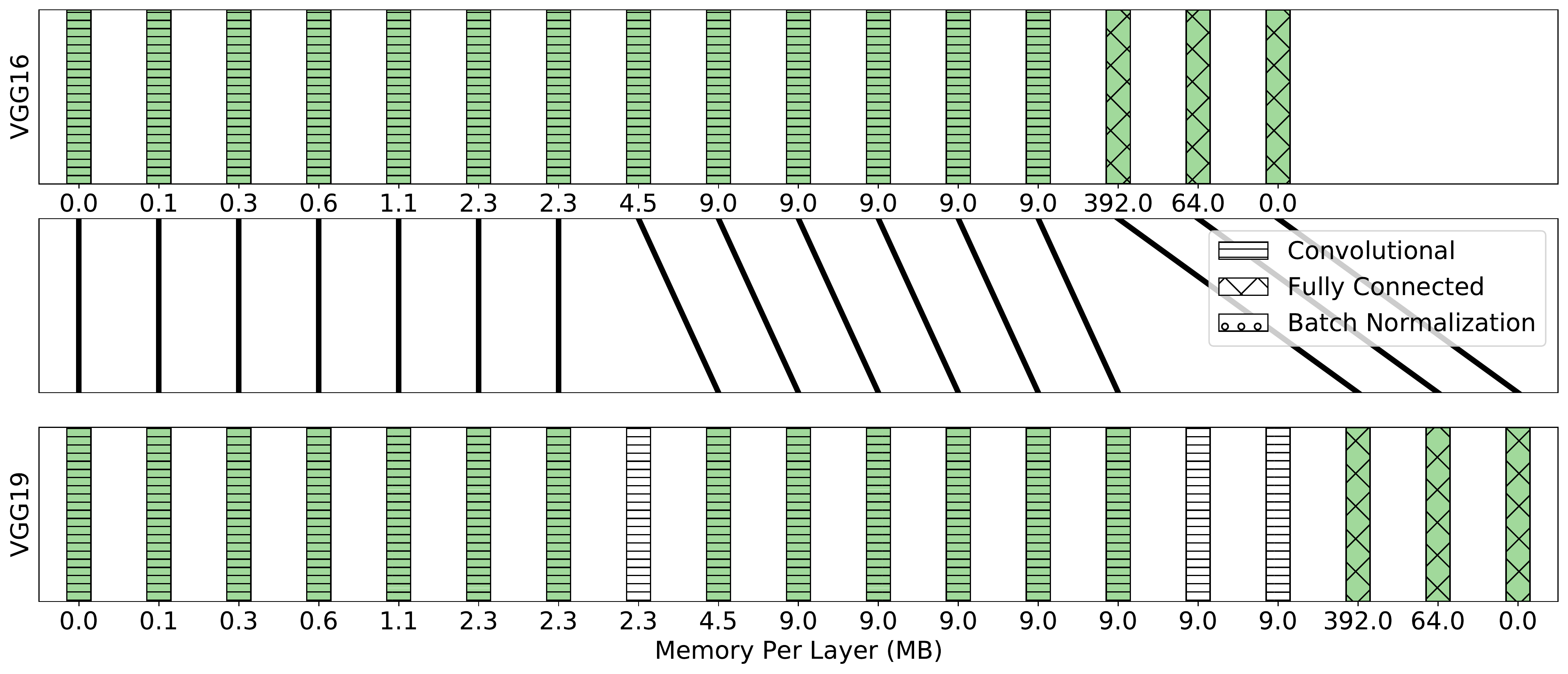}
  \caption{\add{VGG16 and VGG19 are variants within the VGG model family~\cite{vgg}. They share 16/19 layers (13 convolutional and 3 fully-connected). Note that `batch normalization' layers are not present in these models.}}
  \label{fig:deep_dive_sharing1}
\end{figure*}

\begin{figure*}[h!]
  \centering
  \includegraphics[width=0.9\linewidth]{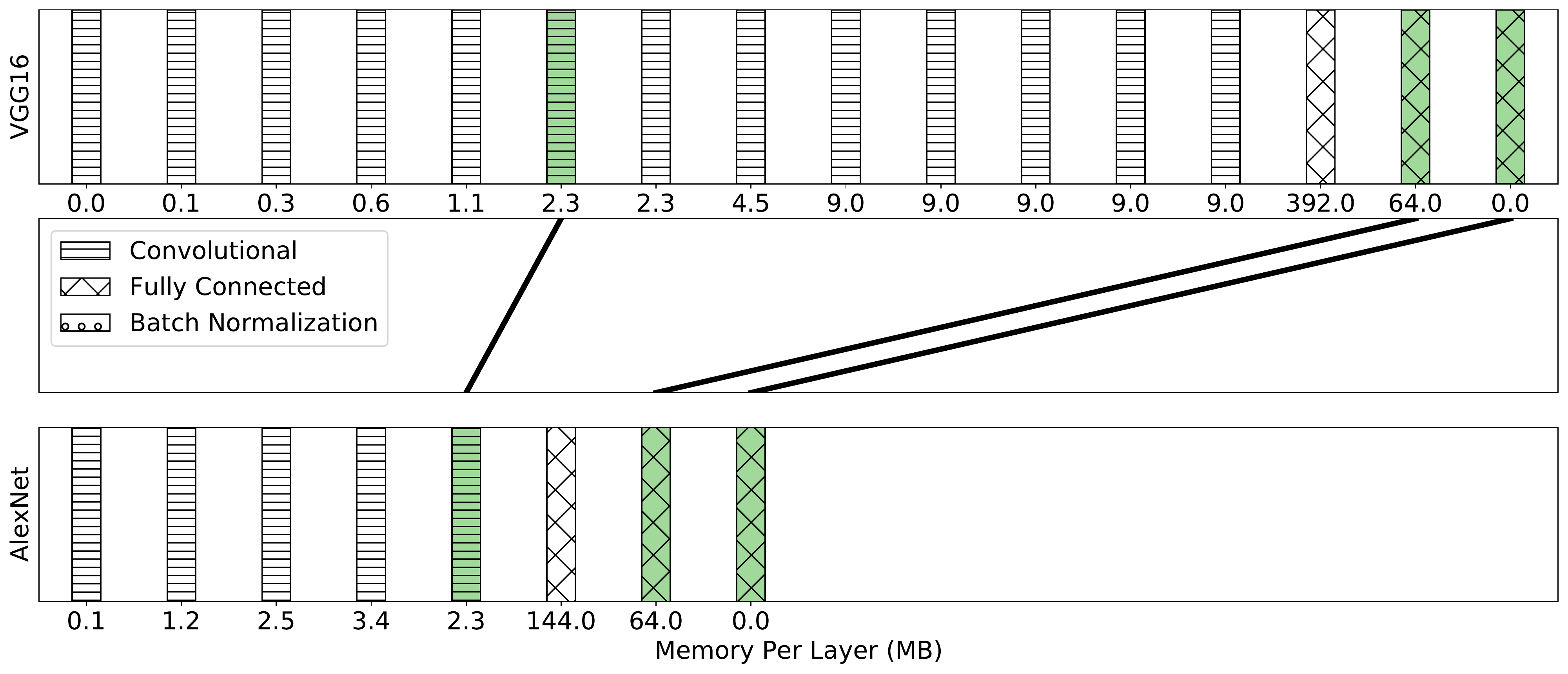}
  \vspace{-5pt}
    \caption{\add{VGG~\cite{vgg} was developed by replacing AlexNet's~\cite{alexnet} large kernels with multiple smaller ones. VGG16 and AlexNet share 3/16 layers (1 convolutional and 2 fully-connected). Note that `batch normalization' layers are not present in these models.}} 
  \label{fig:deep_dive_sharing2}
\end{figure*}

\begin{figure*}[h!]
  \centering
  \includegraphics[width=0.95\linewidth]{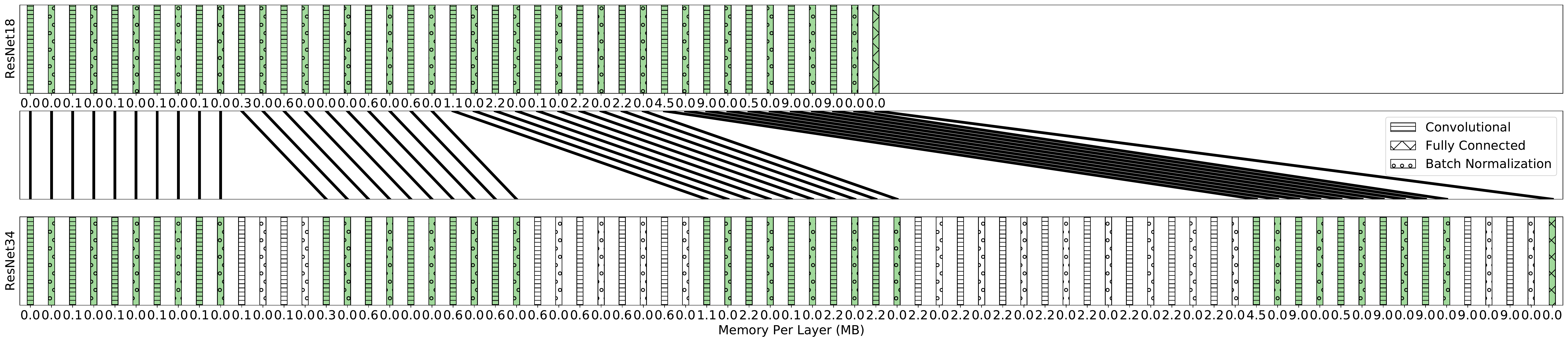}
  \vspace{-5pt}
  \caption{\add{ResNet18 and ResNet34 are variants within the ResNet model family~\cite{resnet}. They share 41/73 layers (20 convolutional, 1 fully-connected and 20 batch normalization).}} 
  \label{fig:deep_dive_sharing3}
\end{figure*}

\begin{figure*}[h!]
   \centering
   \includegraphics[width=\linewidth]{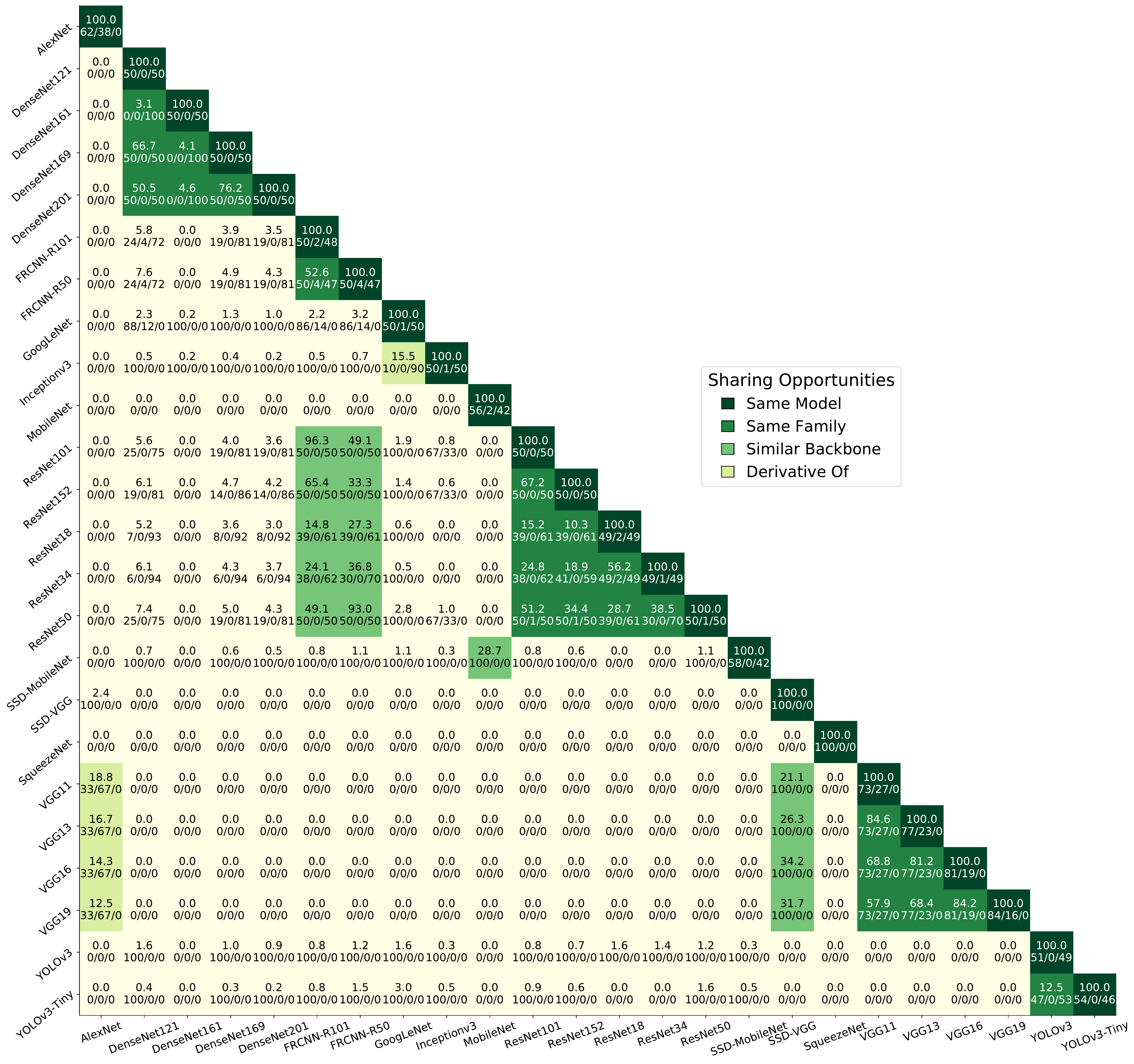}
   \caption{\add{Extended version of Figure~\ref{fig:models_overlap}. For each unique pair of models, we show the percentage of architecturally identical layers and of those layers, the percent breakdown across layer types (\%Convolutional / \%Linear / \%BatchNorm).}}
  \label{fig:model_overlap_complete}
  \end{figure*}

\begin{figure*}[h!]
  \centering
  \includegraphics[width=0.8\linewidth]{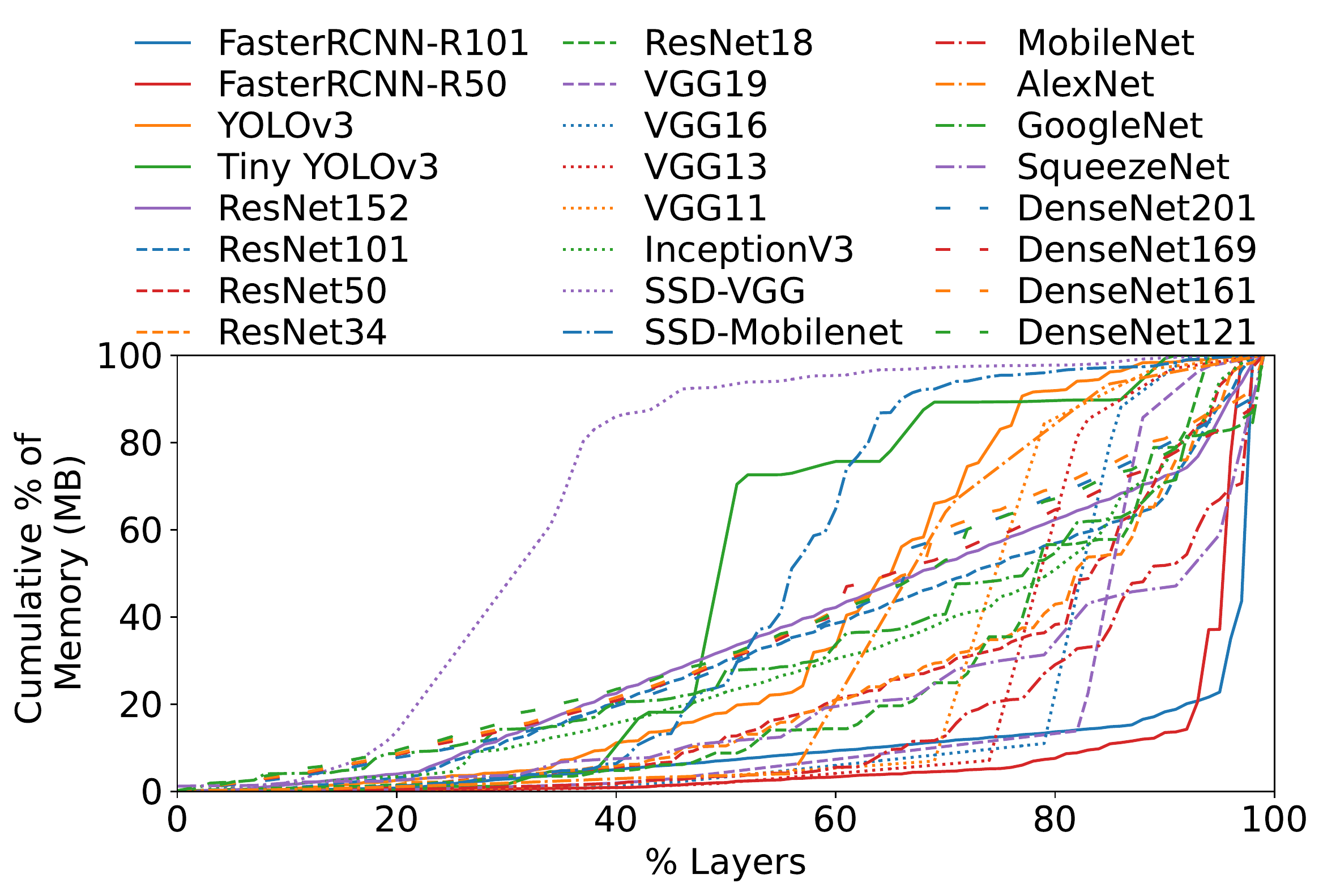}
  \vspace{-10pt}
  \caption{\add{Extended version of Figure~\ref{fig:mem_by_layer_maintext}. Cumulative memory consumed by each model's layer groups moving from start to end of the model.}}
  \label{fig:mem_by_layer}
\end{figure*}

\begin{figure*}[h]
   \raggedright
   \includegraphics[width=\textwidth]{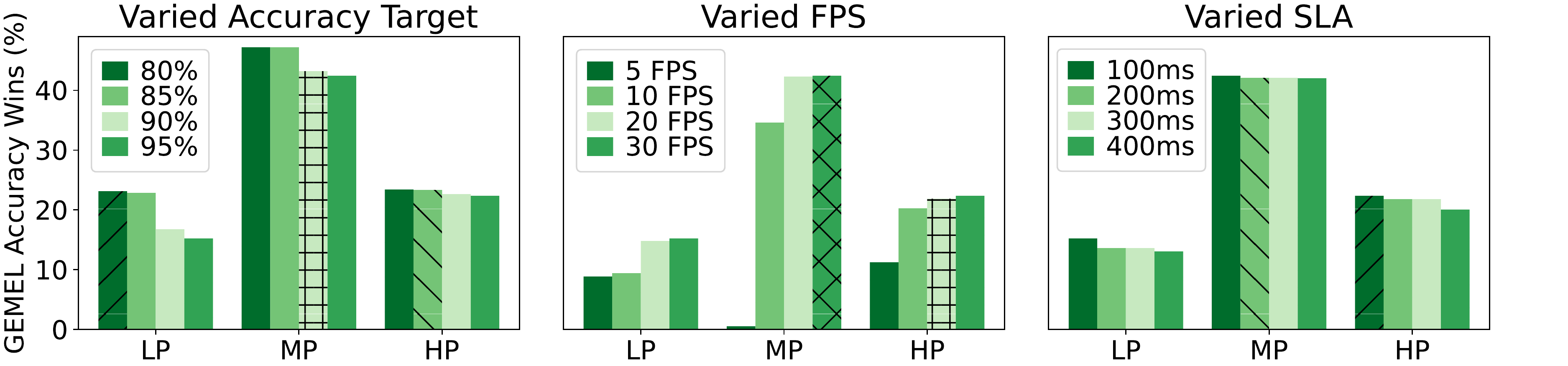}
   \vspace{-10pt}
   \caption{\add{\name{}'s accuracy wins (compared to time/space-sharing alone) with varied accuracy targets, FPS, and SLAs.}}
   \label{fig:vary_params}
\end{figure*}

\begin{figure*}
    \centering
    \includegraphics[width=\textwidth]{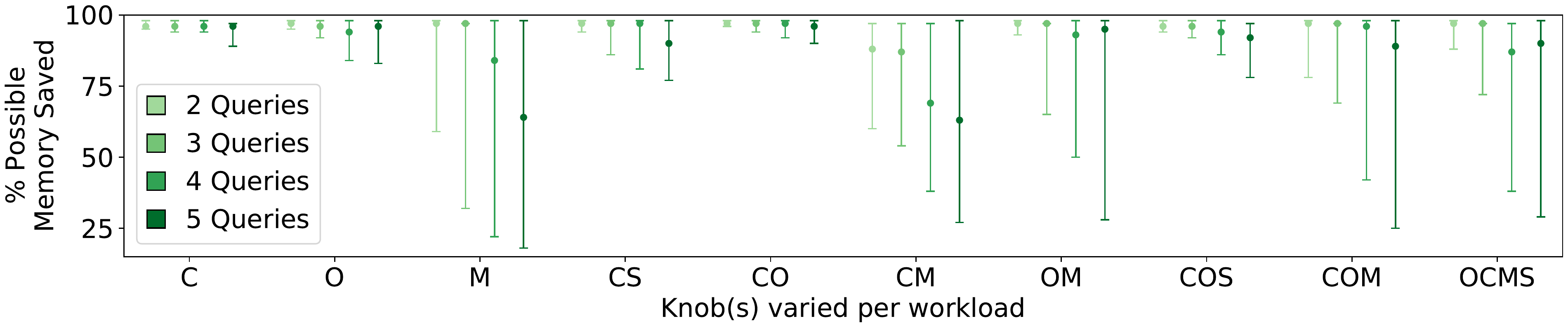}
    \vspace{-5pt}
    \caption{\add{Extended version of Figure~\ref{fig:gen_extensive}. Memory savings across 872 workloads, organized by workload size (color) and knobs varied (x-axis). We plot the median of each distribution (error bars spanning 25-75P). Knobs are labeled as follows: C:Camera, O:Object, M:Model, S:Scene.}}
    \label{fig:gen_extensive_appendix}
\end{figure*}

\begin{figure*}[h!]
   \centering
   \includegraphics[width=\textwidth]{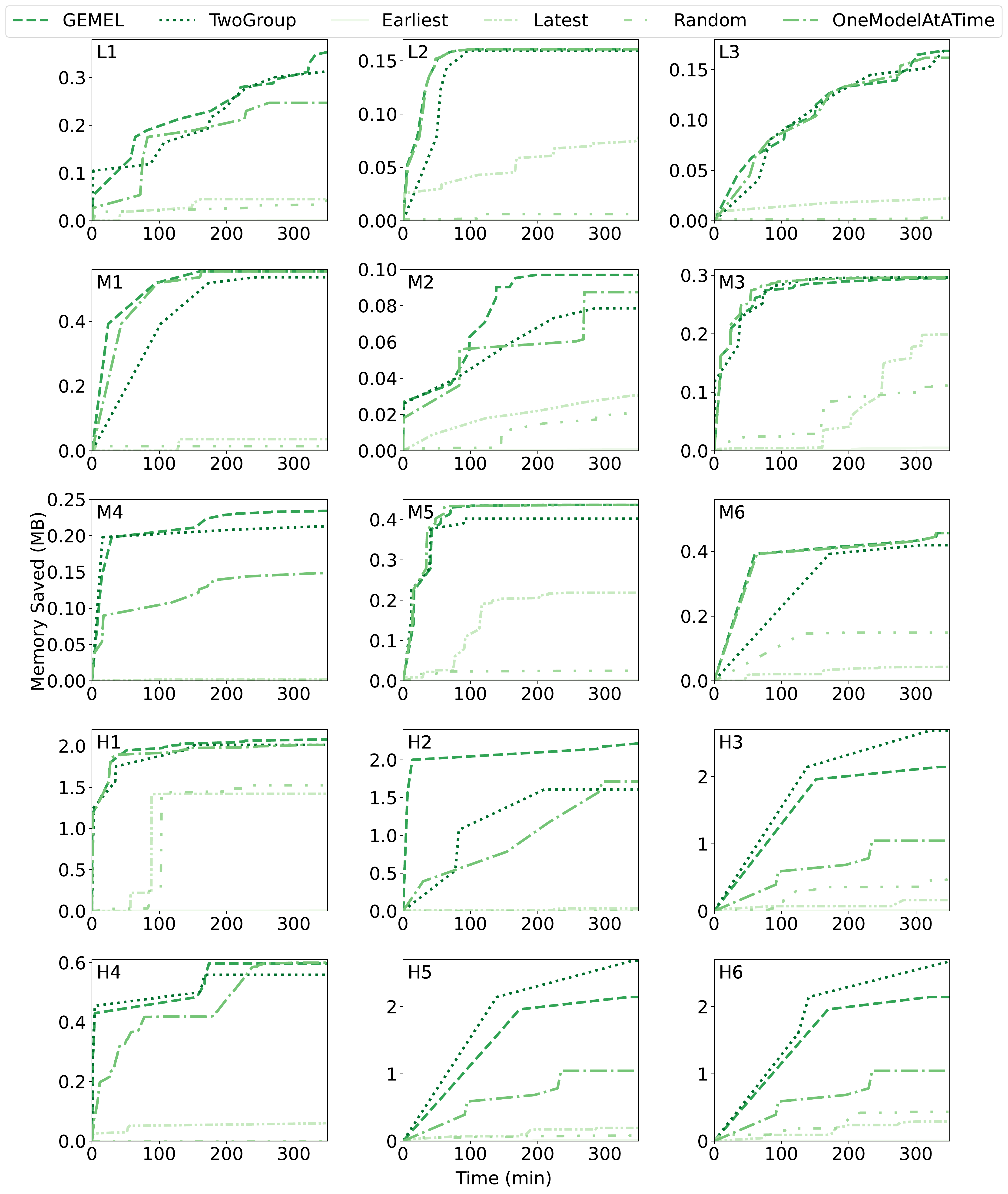}
   \vspace{-10pt}
   \caption{\add{Complete version of Figure~\ref{fig:heuristic_comp}. Comparison of \name{} with other merging heuristics.}}
  \label{fig:compare_heuristics}
  \end{figure*}
  
\end{document}